\documentclass[twocolumn]{autart}    
\usepackage{savesym}
\savesymbol{AND}
\usepackage{graphicx, subfigure}  
                               
\usepackage{amsmath} 
\usepackage{amssymb}  
\usepackage{amsfonts, bm}
\usepackage{algorithm,algorithmic}
\usepackage{arydshln,leftidx,mathtools,multirow}
\usepackage[dvips]{epsfig}   
\usepackage{psfrag,url}
\usepackage{color}

\renewcommand{\thefigure}{\arabic{figure}}

\newcommand{\algrule}[1][.2pt]{\par\vskip.5\baselineskip\hrule height #1\par\vskip.5\baselineskip}
\newcommand{\algrrule}[1][.1pt]{\par\vskip.5\baselineskip\hrule height #1\par\vskip-.5\baselineskip}

  \usepackage[color]{changebar}
\cbcolor{red}
\begin{document}

\begin{frontmatter}

\title{Error-free approximation of explicit linear MPC through lattice piecewise affine expression\thanksref{footnoteinfo}} 

\thanks[footnoteinfo]{This work was supported in part by the National Natural Science Foundation of China under Grant U1813224, 62173113, and in part by the Science and Technology Innovation Committee of Shenzhen Municipality under Grant GXWD20201230155427003-20200821173613001, JCYJ20200109113412326. The material in this paper was partially presented at the 60th IEEE Conference on Decision and Control, December 13-17, 2021, Austin, Texas, USA.}

\author[Paestum,Rome]{Jun Xu}\ead{xujunqgy@gmail.com},    
\author[Paestum]{Yunjiang Lou}\ead{louyj@hit.edu.cn},               
\author[Baiae]{Bart De Schutter}\ead{b.deschutter@tudelft.nl },  
\author[Berlin]{Zhenhua Xiong}\ead{mexiong@sjtu.edu.cn}

\address[Paestum]{School of Mechanical Engineering and Automation, Harbin Institute of Technology, Shenzhen}  
\address[Rome]{Key Laboratory of System Control and Information Processing, Ministry of Education, Shanghai}             
\address[Baiae]{Delft Center for Systems and Control, Delft University of Technology, The Netherlands}        
\address[Berlin]{State Key Laboratory of Mechanical Systems and Vibration, Shanghai Jiao Tong University, Shanghai, China}


\begin{abstract}                          
In this paper, the disjunctive and conjunctive lattice piecewise affine (PWA) approximations of explicit linear model predictive control (MPC) are proposed. The training data are generated uniformly in the domain of interest, consisting of the state samples and corresponding affine control laws, based on which the lattice PWA approximations are constructed. Re-sampling of data is also proposed to guarantee that the lattice PWA approximations  are identical to the explicit MPC control law in the unique order (UO) regions containing the sample points as interior points. \cbstart Additionally, if all the distinct affine functions have been sampled, the disjunctive lattice PWA approximation constitutes a lower bound while the conjunctive lattice PWA approximation formulates an upper bound of the original optimal control law. \cbend The equivalence of the two lattice PWA approximations then guarantees that the approximations are error-free in the domain of interest, which is tested through a statistical guarantee. The complexity of the entire procedure is analyzed, which is polynomial with respect to the number of samples. The performance of the proposed approximation strategy is tested through two simulation examples, and the results show that with a moderate number of sample points, we can construct lattice PWA approximations that are equivalent to the optimal control law of the explicit linear MPC.
\end{abstract}

\end{frontmatter}

\section{Introduction}
Model predictive control (MPC) is currently the most popular control methodology employed in process control and its impact on industry has been recognized widely \cite{Samad2017survey}.  In MPC, the control action is calculated through solving a finite-horizon open-loop optimal control problem at each sampling instant, which is computationally expensive and only suitable for systems with slow dynamics. Therefore, for fast dynamical systems, in order to use MPC, the complexity of the online optimization should be reduced. A natural thought is to move the online optimization offline, which is the idea of explicit MPC. Explicit MPC was first introduced in \cite{Bemporad2002explicit}, in which the linear MPC problem is formulated as a multi-parametric quadratic programming (mpQP) problem and solved offline. The optimal control law is proved to be continuous piecewise affine (PWA) with respect to the state. The subregions as well as the corresponding affine functions defined on them are recorded. For online implementation, given the current state, one must only find the subregion in which the state lies, and the function evaluation of the corresponding affine function gives rise to the optimal control law. 

However, the offline construction of such subregions, the memory required to store them, and the online search for the right subregion  are the main limitations of explicit MPC \cite{Kvasnica2015region}. Much work has been done to solve these three problems. To overcome the complex offline geometric computations, combinatorial approaches are proposed that are based on implicitly enumerating all possible combinations of active constraints \cite{Gupta2011novel,Moshkenani2018combinatorial}. To reduce the memory required to store the subregions as well as the control laws, region-free explicit MPC is proposed \cite{Kvasnica2015region,Borrelli2010computation}. Moreover, the online search complexity can be reduced by storing additional information \cite{Herceg2013evaluation,Christophersen2007efficient}, \cbstart introducing an improved binary search tree (orthogonal truncated binary search tree) \cite{Bayat2012flexible},  or resorting to the method of convex lifting \cite{Nguyen2017convex,Gulan2019efficient}. \cbend The lattice PWA representation has also been used to {exactly} express the explicit MPC control law, resulting in a much lower storage requirement \cite{Wen2009analytical,Xu2016irredundant}. As the complexity of solving the explicit MPC problem increases exponentially with the size of the optimization problem, all these methods can only alleviate the computational burden to some extent.

Another idea is to formulate the approximate MPC controller \cite{Bemporad2001suboptimal,Bemporad2011ultra} or semi-explicit MPC controller \cite{Goebel2017semi}. 
 In these methods, training data containing the  values of states and corresponding optimal control laws of the MPC problem are generated, and the approximated controller is constructed using these data. In general, the samples are required to be distributed sufficiently evenly over the  domain \cite{Chakrabarty2016support}. Different approaches have been used to generate the approximation, such as the canonical piecewise linear function \cite{Karg2020efficient}, radial basis functions \cite{CsekHo2015explicit}, wavelets \cite{Summers2011mutlresolution}, and so on. In addition, reinforcement learning has also been used to derive a data-driven MPC control law in \cite{Gros2020data}. In the work of \cite{Bemporad2011ultra}, \cite{Summers2011mutlresolution}, and \cite{Scibilia2009approximate}, the approximations are based  on particular partitions of the domain of interest, and the interpolation based algorithm can be developed \cite{Pavlov2020minimax}. In fact, the partitions of the domain of interest employed in these works are different from the domain partitions in the explicit linear MPC law, e.g.,  
 the continuous PWA solution of the linear MPC problem has not been fully explored in the approximation. 

\cbstart To resemble explicit MPC control law to a larger extent, the information of the local affine functions of the linear explicit MPC is utilized to derive the lattice PWA approximation. In our previous work, the lattice PWA \emph{representation} of the explicit MPC control law was derived, which however also scales poorly with the dimension of the parameters.  In this paper, in order to handle more complex problems, we present an \emph{error-free approximation} that consists of disjunctive and conjunctive lattice PWA approximations, which coincide with the explicit MPC control in a statistical sense. Moreover, the offline calculation complexity depends mainly on the number of sample points, which scales well with the state dimension. \cbend 
 A preliminary thought of the disjunctive lattice PWA approximation of explicit linear MPC was presented in \cite{Xu2021lattice}, in which the approximated control law is not guaranteed to be error-free. However, in this work, under mild assumptions, the equivalence of the disjunctive and conjunctive approximations guarantees that the two approximations are identical to the optimal control law in the domain of interest. The approximation can also be simplified to further lower the storage requirements and online computational complexity. 

The rest of this paper is organized as follows. Section \ref{sec:pre} gives the preliminaries about the  explicit linear MPC problem and the lattice PWA expression. The offline approximations of the explicit linear MPC control law through the lattice PWA expression are given in detail in Section 3, in which the sampling and re-sampling procedures,  as well as the simplification of the approximation are provided. In Section 4, the approximation error and the complexity of the proposed procedure are analyzed. Section 5 provides the simulation results and the paper ends with conclusions and plans for future work in Section 6. 


\section{Preliminaries}\label{sec:pre}

\subsection{Explicit linear MPC problem}\label{sec:explicit_mpc}
In particular, MPC for a discrete-time linear time-invariant system can be cast as the following optimization problem at time step $t$:

\begin{subequations}\label{optimproblem}
\begin{align}
\min\limits_{U}   & \Bigg\{J(U,\bm x_0)= v_{N_p}(\bm x_{N_p})+\sum\limits_{k=0}^{N_p-1}v(\bm x_k, \bm u_k) \Bigg\}\\
 \mbox{s.t.}~   & \bm x_{k+1}=A\bm x_k+B\bm u_k, k=0, \ldots, N_p-1\label{linear_system}\\
 {}&(\bm x_k, \bm u_k) \in \mathcal{G}, k=0, \ldots, N_p-1\\
 {}&\bm x_{N_p} \in \mathcal{F}
\end{align}
\end{subequations}
in which the optimized variable is $U=[\bm u_0^T, \ldots, \bm u_{N_p-1}^T]^T$, $N_p$ is the prediction horizon, the variables $\bm x_{k} \in \mathbb{R}^{n_x}$ and $\bm u_k \in \mathbb{R}^{n_u}$ denote the predicted state and  input at time step $k$, respectively, using (\ref{linear_system}).  The terminal penalty is denoted as $v_N$ and $v(\cdot, \cdot)$ is the stage cost; $\mathcal{G}$ and $\mathcal{F}$ are full-dimensional polyhedral sets of appropriate dimensions. In this paper, we assume a strictly convex cost, i.e., $v_N=\bm x_N^T Q_N \bm x_N, v(\bm x_k, \bm u_k)=\bm x_k^TQ_k \bm x_k+\bm u_k^TQ_u\bm u_k$, in which  $Q_u \succ 0, Q_k, Q_N\succeq 0$.
After solving the optimization problem (\ref{optimproblem}), the optimal $U^*=[({\bm u_0^*})^T, \ldots, ({\bm u_{N_p-1}^*)}^T]^T$ is obtained, and only $\bm u_0^*$ is applied to the system. The optimization problem is subsequently reformulated and solved at the next time steps $t=1,2,\ldots$ by updating the given state vector $\bm x_0$.

It has been proved in \cite{Bemporad2002explicit} that the solution $U^*$ is a \emph{\textbf{continuous PWA function}} of the state $\bm x_0$, and we use $\bm x$ instead hereafter in this paper.
In fact, this conclusion is obtained through solving an mpQP problem of the form
\begin{equation}\label{mp-qp}
\begin{array}{rl}
\min\limits_U & \frac{1}{2}U^THU+\bm x^T FU\\
s.t. & GU \leq \bm w+E\bm x
\end{array}
\end{equation}
where $U \in \mathbb{R}^{N_p\cdot n_u}$ is the vector of optimization variables, the parameter vector is $\bm x \in \mathbb{R}^{n_x}$, and the matrices $H, F, G$, and $E$ are calculated through the optimization problem (\ref{optimproblem}) \cite{Bemporad2002explicit}. Under the assumption that  $Q_k,Q_N \succeq 0, Q_u\succ 0$, we have $H \succ 0$.

The definition of a continuous PWA function as well as the lemma concerning the continuous PWA property of the solution to the mpQP problem is presented as follows.

\begin{defn}\label{def_cpwa}\cite{Chua1988} 
A function $f:\Omega  \rightarrow \mathbb{R}^m$, where $\Omega \subseteq \mathbb{R}^{n_x}$ is convex, is said to be continuous PWA if it is continuous on the domain $\Omega$ and the following conditions are satisfied:
\begin{enumerate}
\item The domain space $\Omega $ is divided into a finite number of nonempty convex polyhedra, i.e., $\Omega=\cup_{i=1}^{\hat{N}} \Omega_{i},~\Omega_i \neq \emptyset$, the polyhedra are closed and have non-overlapping interiors, $\mathrm{int}(\Omega_i) \cap \mathrm{int}(\Omega_j) = \emptyset, ~\forall i,j \in \{1,\ldots,\hat{N}\}, i \neq j$. These polyhedra are also called local regions. The boundaries of the polyhedra  are nonempty sets in ($n-1$)-dimensional space.
\item In each local region $\Omega_i$, $f$ equals a local affine function $u_{\mathrm{loc}(i)}$:
\begin{equation*}
f(\bm x)=u_{\mathrm{loc}(i)}(\bm x), ~\forall x \in \Omega_i.
\end{equation*}
 \end{enumerate}
\end{defn}

\begin{lem}\label{lem:mpc_cpwa}\cite{Bemporad2002explicit}
Considering the mpQP problem (\ref{mp-qp}) and assuming that $H \succ 0$, then the set of feasible parameters $\Omega \subset \mathbb{R}^{n_x}$ is convex, the optimizer $U^* : \Omega \rightarrow \mathbb{R}^{p\cdot n_u}$ is continuous PWA, and the value function $J^*: \Omega \rightarrow \mathbb{R}$ is continuous convex and piecewise quadratic.
\end{lem}

The details of constructing such continuous PWA function are given as follows. 

First, the mpQP problem can be rewritten in the form
\begin{equation}\label{mp-qp2}
\begin{array}{rl}
\min\limits_{\bm z} & \frac{1}{2}\bm z^T H \bm z\\
s.t. & G\bm z \leq \bm w+S\bm x
\end{array}
\end{equation}
by letting $\bm z=U+H^{-1}F^T\bm x$ and $S=E+GH^{-1}F^T$. Once the optimal solution $\bm z^*$ of the optimization problem (\ref{mp-qp2}) is available, we can easily obtain the optimal $U^*$ as
\[
U^*=\bm z^*-H^{-1}F^T\bm x.
\]

\cbstart The optimal solution $\bm z^*$ for a fixed $\bm x$ is fully characterized by the Karush-Kuhn-Tucker (KKT) conditions:\cbend
\begin{subequations}\label{eq:KKT}
\begin{align}
&H\bm z^*+G_{\mathcal{A}^*}^T\bm \lambda^*+G_{\mathcal{N}^*}^T\bm \mu^*=0 \label{first_order_condition}\\
&G_{\mathcal{A}^*}\bm z^*=\bm w_{\mathcal{A}^*}+S_{\mathcal{A}^*}\bm x \label{active_constraints}\\
&G_{\mathcal{N}^*}\bm z^*<\bm w_{\mathcal{N}^*}+S_{\mathcal{N}^*}\bm x \label{inactive_constraints}\\
 &\bm\lambda^* \geq 0 \label{dual_feasibility}\\
&\bm \mu^*\geq 0\\
&{\bm \lambda^*}^T(G_{\mathcal{A}^*}\bm z^*-\bm w_{\mathcal{A}^*}-S_{\mathcal{A}^*}\bm x)=0\\
&{\bm \mu^*}^T(G_{\mathcal{N}^*}\bm z^*-\bm w_{\mathcal{N}^*}-S_{\mathcal{N}^*}\bm x)=0
\end{align}
\end{subequations}
in which (\ref{active_constraints}) and (\ref{inactive_constraints}) are the active and inactive constraints at $\bm z^*$, respectively. Assuming that $G\in \mathbb{R}^{p\times N_p\cdot n_u}, \bm w\in \mathbb{R}^{p}, S \in \mathbb{R}^{p\times n_x}$, and $G_i, w_i$, and $S_i$ denote the $i$-th row of $G, w$, and $S$, respectively, the active as well as inactive index sets can be written as
\[
\mathcal{A}^*=\{j\in \{1, \ldots,p\}|G_j\bm z^*=\bm w_j+S_j\bm x\}
\]
and
\[
\mathcal{N}^*=\{j \in \{1, \ldots, p\}|G_j \bm z^*<\bm w_j+S_j\bm x\},
\]
respectively. It is apparent that $\mathcal{A}^*=\{1, \ldots, p\} \setminus \mathcal{N}^*$. For an inactive constraint $j$, we have $\bm \mu_j^*=0$. For a particular $\mathcal{A}^*$, and assume $G_{\mathcal{A}^*}$ is full row rank, we have  
\begin{equation}\label{opt_lambda}
\bm \lambda^*=-(G_{\mathcal{A}^*}H^{-1}G_{\mathcal{A}^*}^T)^{-1}(\bm w_{\mathcal{A}^*}+S_{\mathcal{A}^*}\bm x),
\end{equation}
as well as
\begin{equation}\label{opt_z}
\bm z^*=H^{-1}G_{\mathcal{A}^*}^T(G_{\mathcal{A}^*}H^{-1}G_{\mathcal{A}^*}^T)^{-1}(\bm w_{\mathcal{A}^*}+S_{\mathcal{A}^*}\bm x).
\end{equation}
The local region for which the local affine function (\ref{opt_z}) is defined is called \emph{critical region}, and it can be constructed by the constraints of primal feasibility (\ref{inactive_constraints}) and dual feasibility (\ref{dual_feasibility}).

\begin{rem}\label{rem1}
For the case in which $G_{\mathcal{A}^*}$ is not full row rank, i.e., the rows of $G_{\mathcal{A}^*}$ are linearly dependent, the linear independence constraints qualification (LICQ) is violated according to \cite{Nocedal2006numerical}, and this is referred to as primary degeneracy \cite{Bemporad2002explicit} (dual degeneracy cannot  occur as $H \succ 0$). Assuming that the rank of $G_{\mathcal{A}^*}$ is $r$, we can then arbitrarily select $r$ independent constraints, and proceed with the new reduced active index set \cite{Borrelli2003constrained}.
\end{rem}

To search for all the local affine functions and critical regions, one must enumerate all possible active index sets $\mathcal{A}^*$, apply the KKT conditions accordingly, and then the continuous PWA control law can be obtained.
 In the next subsection, the lattice PWA representation is presented, which is used to express the resulting continuous PWA control law
in our previous work \cite{Xu2016irredundant}.

\subsection{Lattice PWA representation}\label{sec:lpwa_representation}
It is stated in \cite{Xu2016irredundant} that any continuous PWA function can be represented by the lattice PWA representation.

\begin{lem}\label{lem:full_lattice}\cite{Xu2016irredundant}
Letting $f$ be a continuous PWA function defined in Definition 1, then $f$ can be represented as
\begin{equation}\label{eq:ful_lat_representation_d}
f(\bm x)=f_{\mathrm{L,d}}(\bm x)=\max\limits_{i=1,\ldots,N}\min_{j \in I_{\geq,i}}u_j(\bm x), ~\forall x \in \Gamma,
\end{equation}
or
\begin{equation}\label{eq:ful_lat_representation_c}
f(\bm x)=f_{\mathrm{L,c}}(\bm x)=\min\limits_{i=1,\ldots,N}\max_{j \in I_{\leq,i}}u_j(\bm x), ~\forall x \in \Gamma,
\end{equation}
\cbstart in which $I_{\geq,i}=\{j|u_j(\bm x)\geq u_{i}(\bm x),  \forall \bm x \in \Gamma_i\}, I_{\leq, i}=\{j|u_j(\bm x)\leq u_{i}(\bm x), \forall \bm x \in \Gamma_i\}$, and the expressions $\min_{j \in I_{\geq,i}}u_j(\bm x)$ and $\max_{j \in I_{\leq,i}}u_j(\bm x)$ are called terms of $f_{\mathrm{L,d}}$ and $f_{\rm L, c}$, respectively. \cbend The affine function $u_j(\bm x)$ is called a literal. The region $\Gamma_i$ is  a \textbf{unique order (UO) region} that is subset of the local region and the order of the affine functions
\begin{equation}\label{cond_base_region}
u_1(\bm x), \ldots, u_N(\bm x),
\end{equation}
remains unchanged in the interior of $\Gamma_i$.  The expressions (\ref{eq:ful_lat_representation_d}) and (\ref{eq:ful_lat_representation_c}) are called full disjunctive and conjunctive lattice PWA representations, respectively, in which the names ``disjunctive"  and ``conjunctive" come from the terminology in Boolean algebra. 
\end{lem}

Considering a two-dimensional continuous PWA function (\ref{2dfunction}) with four affine pieces, Fig. \ref{fig_ex1} illustrates the UO region and corresponding PWA function.
\begin{exmp}\label{ex0}
\begin{equation}\label{2dfunction}
f=\left\{\begin{array}{ll}
\ell_1(\bm x)=-x_2+1& \mathrm{if}~ \bm x \in \Gamma_1,\\
\ell_2(\bm x)=-x_1+1& \mathrm{if}~ \bm x \in \Gamma_2,\\
\ell_3(\bm x)=x_2+1& \mathrm{if}~ \bm x\in \Gamma_3,\\
\ell_4(\bm x)=x_1+1&\mathrm{if}~ \bm x\in \Gamma_4.
\end{array}\right.
\end{equation}
The polyhedral regions $\Gamma_1, \Gamma_2, \Gamma_3, \Gamma_4$, and the two-dimensional function $f$ are shown in Fig. \ref{fig_ex1}. 
\cbstart
\begin{figure}[htbp]
\centering
   \psfrag{gamma1}[c]{$\Gamma_1$}
     \psfrag{gamma11}[c]{$\Gamma_{11}$}
       \psfrag{gamma12}[c]{$\Gamma_{12}$}
  \psfrag{gamma2}[c]{$\Gamma_2$}
    \psfrag{gamma21}[c]{$\Gamma_{21}$}
      \psfrag{gamma22}[c]{$\Gamma_{22}$}
  \psfrag{gamma3}[c]{$\Gamma_3$}
    \psfrag{gamma31}[c]{$\Gamma_{31}$}
      \psfrag{gamma32}[c]{$\Gamma_{32}$}
  \psfrag{gamma4}[c]{$\Gamma_{4}$}
    \psfrag{gamma41}[c]{$\Gamma_{41}$}
      \psfrag{gamma42}[c]{$\Gamma_{42}$}
  \psfrag{x1}[c]{$x_1$}
  \psfrag{x2}[c]{$x_2$}
    \psfrag{l1}[c]{\color[RGB]{255,0,0}$\ell_1$}
  \psfrag{l2}[c]{\color[RGB]{255,0,0}$\ell_2$}
  \psfrag{l3}[c]{\color[RGB]{255,0,0}$\ell_3$}
    \psfrag{l4}[c]{\color[RGB]{255,0,0}$\ell_4$}
  \psfrag{f}[c]{$f$}
  \subfigure[Function.]{
  \includegraphics[width=0.9\columnwidth]{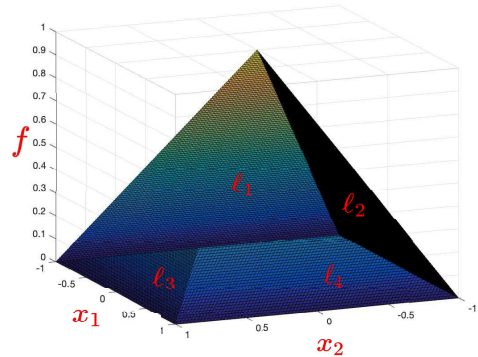}\label{fig_ex1_fun}}\\
  \subfigure[Regions.]{
    \includegraphics[width=0.9\columnwidth]{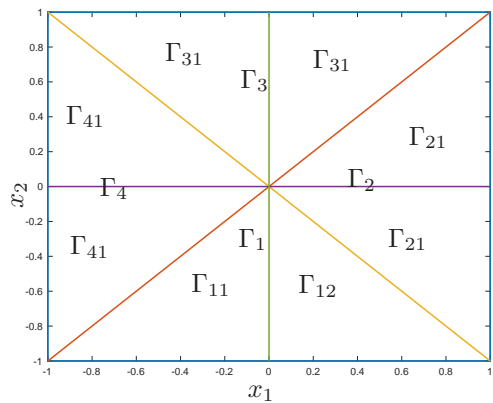}
     \label{fig_ex1_region}}
    \caption{Continuous PWA function in Example \ref{ex0}.}
  \label{fig_ex1}
\end{figure}
\cbend

The regions $\Gamma_1, \Gamma_2, \Gamma_3$, and $\Gamma_4$ are local affine regions, and can be divided into UO regions $\Gamma_{11}, \Gamma_{12}, \ldots, \Gamma_{41}$, and $\Gamma_{42}$. Taking the UO region $\Gamma_{31}$ as an example, the order of affine functions is
\[
\ell_3<\ell_2<\ell_1<\ell_4.
\]
For this continuous PWA function, as it is concave, both the disjunctive and conjunctive lattice PWA representations are
\[
f=\min\{\ell_1, \ell_2, \ell_3, \ell_4\}.
\]
\end{exmp}

According to Lemma \ref{lem:full_lattice}, we can represent a continuous PWA control law using a lattice PWA function (either disjunctive or conjunctive). The disjunctive lattice PWA representation of explicit linear MPC was investigated in \cite{Wen2009analytical} and \cite{Xu2016irredundant}, in which the continuous PWA control law was obtained through the MPT3 toolbox \cite{MPT3} in advance. However, as explained in Section \ref{sec:explicit_mpc}, for problems with a large number of constraints and a high-dimensional state, the number of possible combinations of active constraints increases exponentially and the derivation of the explicit MPC solution  is extremely computationally expensive.  Hence, in this paper, we propose an approximated continuous PWA control law by sampling only a set of states in the domain of interest. We show that this approximation utilizes the local affine property of the original explicit MPC control law and is identical to the original control law at the sample points  and in the UO regions the sample points lie. In addition, under mild assumptions, the lattice PWA approximations are identical to the explicit MPC control law in the  domain of interest.

\section{Lattice PWA approximation of explicit linear MPC control law}
\subsection{Generation of sample points in the interior of UO regions}\label{sec:generate_points}

\cbstart
As indicated in Lemma \ref{lem:mpc_cpwa}, the explicit linear MPC control law $U^*$ is a continuous PWA function with respect to the state $\bm x$. Then the first element of $U^*$, which is $\bm u_0^*$, is also a continuous PWA function of $\bm x$, i.e., $\bm u_0^*$ is affine in the local regions that $\bm x$ lies in. For simplicity, we omit the subscript in $\bm u_0^*$ hereafter in the paper, and use  $\bm u^*$ instead. The sample points $(\bm x_i,  \bm u_i(\bm x_i)) \in \mathcal{X}_1 \times \mathcal{U}_1$ are generated in the domain of feasible parameters, in which $\bm u_i(\bm x)$ is the affine function at $\bm x_i$ such that
\[
\bm u_i(\bm x_i)=\bm u^*(\bm x_i).
\]

For simplicity, we consider the case $n_u=1$; note however that the proposed methodology can be easily extended to the case when $n_u>1$. Moreover, the domain of interest is assumed to be a hyperbox.
\cbend

In this subsection, the training points  $\bm x_i \in \mathcal{X}_1$  are required to be in the interior of UO regions, i.e., $\bm x_i \in \mathrm{int}(\Gamma(\bm x_i))$, in which $\Gamma(\bm x_i)$ is the corresponding UO region. This means that if  $u_{j_1}(\bm x_i)>u_{j_2}(\bm x_i)$ holds, then we have
\[
u_{j_1}(\bm x)>u_{j_2}(\bm x), \forall \bm x \in \Gamma(\bm x_i).
\]
\cbstart
As a matter of fact, if there are no affine functions $u_{j_1}(\bm x)$ and $u_{j_2}(\bm x)$ such that $u_{j_1}(\bm x_i)=u_{j_2}(\bm x_i)$, then $\bm x_i \in \mathrm{int}(\Gamma(\bm x_i))$. If there  is a point $\bm x_i$ that is not in the interior of the corresponding UO region, i.e., there exist $j_1, j_2$, such that $u_{j_1}(\bm x_i)=u_{j_2}(\bm x_i)$, then a perturbation should be applied, i.e., $\bm x_i=\bm x_i+\delta$, such that $\bm x_i$ is in the interior of the UO region. This will be shown in Example \ref{ex1} in Section 3.3. It will be shown in 
Lemma \ref{lem:lattice_approx} that the condition that $\bm x_i$ is in the interior of UO regions is required for the equivalence of lattice PWA approximations and the original optimal control law in UO regions.

It should be noted that for the checking of UO regions, only the information of sampled affine functions is needed, and the global information of the PWA function is unknown, e.g., whether there is some $u_j$ that has not been sampled or how these sampled affine functions are connected. 
\cbend


Algorithm \ref{alg:sampling} describes the sampling of training points. 

\algrrule[0.8pt]
\begin{alg}
{Sampling of training points in explicit linear MPC control law.}
\label{alg:sampling}
\algrule[0.5pt]
\begin{algorithmic}[1]
\hspace{-4ex} \textbf{Input:} Linear MPC problem, the number of sample points $N_1$, sample domain $\Omega$.\\
\hspace{-4ex} \textbf{Output:} Sample data set $\mathcal{X}_1 \times \mathcal{U}_1$.
\STATE $\mathcal{X}_1=\emptyset$, $\mathcal{U}_1=\emptyset$. 
\FOR{$i=1$ to $N_1$}
\STATE Generate a grid point $\bm x_i$ in $\Omega$.
\IF {$\bm x_i$ is not an interior point of some UO region}
\STATE Apply a perturbation until $\bm x_i$ is an interior point of $\Gamma(\bm x_i)$.\label{line:perturbation}
\ENDIF
\STATE Solve corresponding QP problem (\ref{mp-qp2}) by letting $\bm x=\bm x_i$.\label{line:qp}
\STATE Determine active and inactive index sets $\mathcal{A}^*$ and $\mathcal{N}^*$, respectively.
\STATE Solve KKT conditions (\ref{eq:KKT}) to obtain the affine function $\bm z_i^*$ through (\ref{opt_z}). 
\STATE Obtain optimal input, i.e., $U_i(\bm x_i)$ and $u_i(\bm x_i)$.
\ENDFOR
\algrule[0.5pt]
\end{algorithmic}
\end{alg}

For a feasible state $\bm x_i$, line \ref{line:qp} in Algorithm \ref{alg:sampling} states that the optimal $\bm z_i^*=U_i^*+H^{-1}F^T\bm x_i$ can be obtained through the QP problem (\ref{mp-qp2}), which, together with the information of $\bm x_i$, determines the active and inactive constraints (\ref{active_constraints}) and (\ref{inactive_constraints}), respectively. Therefore, the index set $\mathcal{A}_i^*$ as well as $\mathcal{N}_i^*$ is fixed, and if the matrix $G_{\mathcal{A}_i^*}$ is full row rank, the affine function $\bm z_i(\bm x_i)$ can be calculated through (\ref{opt_z}) (the rank-deficient case can be handled as indicated in Remark \ref{rem1}). Then we have
\begin{equation}\label{U_affine}
U_i(\bm x_i)=\bm z_i^*(\bm x_i)-H^{-1}F^T\bm x_i,
\end{equation}
and 
\begin{equation}\label{uU_transform}
 u_i(\bm x_i)=\left[\begin{array}{cccc}
\mathbf{I}_{n_u \times n_u}&\bm 0&\cdots&\bm 0
\end{array}\right]U_i(\bm x_i)
\end{equation}
in which $\mathbf{I}_{n_u\times n_u}$ is the identity matrix with size $n_u \times n_u$.

After evaluating Algorithm \ref{alg:sampling}, we can  obtain the sample dataset $\mathcal{X}_1 \times \mathcal{U}_1$, in which $\mathcal{U}_1$ is a set of affine functions $u_i(\bm x_i)$. It is noted that compared with ordinary sampling, in which only the evaluation of $u_i^*(\bm x_i)$ is available, here the corresponding affine function is also recorded, which can be used for the lattice PWA approximation in Section \ref{sec:lattice_sample}.

\subsection{Lattice PWA approximation based on sample points}\label{sec:lattice_sample}
We now  derive both the disjunctive and conjunctive lattice PWA approximations based on the  sample dateset $\mathcal{X}_1 \times \mathcal{U}_1$.

The disjunctive lattice PWA approximation is constructed via the sample points and can be expressed as follows:
\begin{equation}\label{eq:lattice_approximation_d}
\hat{f}_{\mathrm{L, d}}(\bm x)=\max\limits_{i=1,\ldots,N_1}\min_{j \in J_{\geq,i}}u_j(\bm x),
\end{equation}
in which the index set $J_{\geq,i}$ is described as
\begin{equation}\label{eq:index_Jgeq}
J_{\geq,i}=\{j|u_j(\bm x_i)\geq u_i(\bm x_i)\}.
\end{equation}

Similarly, the conjunctive lattice PWA approximation can be described as follows:
\begin{equation}\label{eq:lattice_approximation_c}
\hat{f}_{\mathrm{L, c}}(\bm x)=\min\limits_{i=1,\ldots,N_1}\max_{j \in J_{\leq,i}}u_j(\bm x),
\end{equation}
in which the index set $J_{\leq,i}$ is described as
\begin{equation}\label{eq:index_Jleq}
J_{\leq,i}=\{j|u_j(\bm x_i)\leq u_i(\bm x_i)\}.
\end{equation}

Compared with the full disjunctive and conjunctive lattice PWA representations (\ref{eq:ful_lat_representation_d}) and (\ref{eq:ful_lat_representation_c}), respectively, we can see that the lattice PWA approximations (\ref{eq:lattice_approximation_d}) and (\ref{eq:lattice_approximation_c}) only consider the order of local affine control laws at each sample point. Under certain conditions as shown in Assumption \ref{assump0}, the lattice PWA approximations (\ref{eq:lattice_approximation_d}) and (\ref{eq:lattice_approximation_c}) coincide with the explicit linear MPC control law at the sample points.

\begin{assum}\label{assump0}
The terms in both disjunctive and conjunctive lattice PWA approximations satisfy,
\begin{equation}\label{eq:assump_d}
\min\limits_{j \in J_{\geq, i}}u_j(\bm x_k)\leq u_k(\bm x_k), \forall i, k \in \{1, \ldots, N_1\}
\end{equation}
and
\begin{equation}\label{eq:assump_c}
\max\limits_{j \in J_{\leq, i}}u_j(\bm x_k)\geq u_k(\bm x_k), \forall i, k \in \{1, \ldots, N_1\}
\end{equation}
\end{assum}

\begin{lem}\label{lem:lattice_approx}
Assume that the disjunctive and conjunctive lattice PWA approximations are constructed through (\ref{eq:lattice_approximation_d}) and (\ref{eq:lattice_approximation_c}), respectively. Supposing that Assumption \ref{assump0} holds, 
 then we have
\begin{equation}\label{lattice==pwa}
\hat{f}_{\mathrm{L, d}}(\bm x)=u^*(\bm x), \forall \bm x \in \Gamma(\bm x_i), \forall \bm x_i\in \mathcal{X}_1
\end{equation}
and
\begin{equation}\label{eq:lattice==pwa2}
\hat{f}_{\mathrm{L, c}}(\bm x)=u^*(\bm x), \forall \bm x \in \Gamma(\bm x_i), \forall \bm x_i\in \mathcal{X}_1.
\end{equation}
\end{lem}
\begin{pf}

We first prove (\ref{lattice==pwa}) for the disjunctive case.

As all the sample points $\bm x_i$ are \emph{interior points for some UO regions}, according to (\ref{eq:assump_d}), we have
\[
\min\limits_{j \in J_{\geq, k}}u_j(\bm x) \leq u_i(\bm x), \forall i, k \in \{1, \ldots, N_1\}, \forall \bm x \in \Gamma(\bm x_i)
\]

Besides, the equality holds for $i=k$, i.e.,
\[
\min\limits_{j \in J_{\geq, i}}u_j(\bm x)=u_i(\bm x), \forall i\in \{1, \ldots, N_1\}, \forall \bm x\in \Gamma(\bm x_i).
\]
Combining with the disjunctive lattice PWA approximation (\ref{eq:lattice_approximation_d}), we have
\[
\hat{f}_{\rm L, d}(\bm x) = u_i(\bm x), \forall k \in \{1, \ldots, N_1\}, \forall \bm x \in \Gamma(\bm x_i).
\]

As $u^*(\bm x)= u_i(\bm x), \forall i \in \{1, \ldots, N_1\}, \forall \bm x \in \Gamma(\bm x_i)$, we then have
 (\ref{lattice==pwa}).

The conjunctive case can be proved similarly.

\end{pf}

\begin{rem}
It is noted that the lattice PWA approximation differs from the other approximations in that the lattice PWA approximation equals the original control law not only at the sample points, but also in the UO regions containing the sample points as interior points, as (\ref{lattice==pwa}) and (\ref{eq:lattice==pwa2}) show. To achieve this, for each sample point, not only is the value of the corresponding control law recorded, but the specific affine expression is as well, as Algorithm \ref{alg:sampling} shows.
\end{rem}

\subsection{Re-sampling}

In general, when we generate a moderate number of sample points, (\ref{eq:assump_d}) and (\ref{eq:assump_c}) hold. However, there are situations when the two inequalities are not valid, the following gives the resampling method such that both (\ref{eq:assump_d}) and (\ref{eq:assump_c}) hold.


%


\subsubsection{Guaranteeing the validity of Assumption \ref{assump0}}\label{sec:guarantee1}


Taking the disjunctive lattice PWA approximation as an example, if (\ref{eq:assump_d}) is violated  for some $\bm x_{\alpha}, \bm x_{\beta} \in \mathcal{X}_1$, i.e.,
\begin{equation}\label{cond_violate}
\min\limits_{j \in J_{\geq, \alpha}}u_j(\bm x_{\beta})>u^*(\bm x_{\beta})=u_{\beta}(\bm x_{\beta}),
\end{equation}
we can add sample points in the line segment  
\begin{equation}\label{def:line-segment}
\mathcal{L}(\bm x_{\alpha}, \bm x_{\beta}) =\lambda \bm x_{\alpha}+(1-\lambda) \bm x_{\beta}, \lambda \in (0,1)
\end{equation}
such that (\ref{eq:assump_d})  is satisfied for $\alpha$ and $\bm x_{\beta}$, which is shown in Lemma \ref{lem_find_points}.

\begin{lem}\label{lem_find_points}
Assuming that there are two points $\bm x_{\alpha}$ and $\bm x_{\beta}$ such that (\ref{cond_violate}) holds, then there must be some point $\bm x_{\gamma}\in \mathcal{L}(\bm x_{\alpha}, \bm x_{\beta})$, which is defined in (\ref{def:line-segment}), and the corresponding control solution $u_{\gamma}$, such that the inequality
\begin{equation}\label{concl_lem8}
u_{\gamma}(\bm x_{\alpha})\geq u_{\alpha}(\bm x_{\alpha}), u_{\gamma}(\bm x_{\beta}) \leq u_{\beta}(\bm x_{\beta}),
\end{equation}
holds.

Furthermore, by adding $\bm x_{\gamma}$ to the sample dataset, we have
\begin{equation}\label{eq:concl-lem8.2}
\min\limits_{j \in J_{\geq, \alpha}}u_j(\bm x_{\beta})\leq u_{\beta}(\bm x_{\beta}).
\end{equation}

\end{lem}

\begin{pf}
As the optimal control solution $u^*$ is continuous PWA, it is still continuous PWA when restricted to the line segment $\mathcal{L}(\bm x_{\alpha}, \bm x_{\beta})$.

 Defining an index set $\mathrm{aff}(\bm x_{\alpha}, \bm x_{\beta})$ as
 \[
 \mathrm{aff}{(\bm x_{\alpha}, \bm x_{\beta})}=\{j| \exists \bm x \in \mathcal{L}(\bm x_{\alpha}, \bm x_{\beta})~\mbox{such that}~u^*(\bm x)=u_j(\bm x)\},
 \]
 i.e., the index set $\mathrm{aff}{(\bm x_{\alpha}, \bm x_{\beta})}$ includes all the indices of affine functions in $u^*(x)$ when restricted to $\mathcal{L}(\bm x_{\alpha}, \bm x_{\beta})$.
 
According to \cite{Xu2016irredundant}, we have
\[
\min\limits_{j \in S_{\geq,\alpha}}u_j (\bm x) \leq u_{\beta}(\bm x_{\beta}), \forall \bm x \in \mathcal{L}(\bm x_{\alpha}, \bm x_{\beta}), 
\]
in which $S_{\geq,\alpha}$ is the index set such that
\[
S_{\geq,\alpha}=\{j \in \mathrm{aff}{(\bm x_{\alpha}, \bm x_{\beta})}|u_j(\bm x_{\alpha}) \geq u_{\alpha}(\bm x_{\alpha})\}.
\]
Apparently, there should be some $\gamma \in S_{\geq,{\alpha}}$, such that (\ref{concl_lem8}) is valid.

Therefore, if we add one of these $\bm x_{\gamma}$ to the sample point set, we have
\[
\gamma \in S_{\geq, \alpha} \subset J_{\geq, \alpha}.
\]
Hence (\ref{eq:concl-lem8.2}) is valid.
\end{pf}


It is noted that for the case when (\ref{eq:assump_c}) is violated, we have similar results.

\begin{cor}
Assuming that there are two points $\bm x_{\alpha}$ and $\bm x_{\beta}$ such that the following inequality holds:
\begin{equation}\label{eq:violate2}
\max\limits_{j \in J_{\leq, \alpha}}u_j(\bm x_{\beta})<u^*(\bm x_{\beta})=u_{\beta}(\bm x_{\beta}),
\end{equation}
then there must be some point $\bm x_{\gamma}\in \mathcal{L}(\bm x_{\alpha}, \bm x_{\beta})$, and the corresponding control solution $u_{\gamma}$ satisfies the following inequality:
\begin{equation}\label{eq:concl_cor}
u_{\gamma}(\bm x_{\alpha})\leq u_{\alpha}(\bm x_{\alpha}), u_{\gamma}(\bm x_{\beta}) \geq u_{\beta}(\bm x_{\beta}).
\end{equation}

Additionally, by adding $\bm x_{\gamma}$ to the sample dataset, we have
\begin{equation}\label{eq:concl_cor2}
\max\limits_{j \in J_{\leq, \alpha}}u_j(\bm x_{\beta})\geq u_{\beta}(\bm x_{\beta}).
\end{equation}
\end{cor}


When (\ref{cond_violate}) or (\ref{eq:violate2}) holds, in order to construct  a continuous PWA function and to ensure the validity of (\ref{eq:assump_d}) and (\ref{eq:assump_c}) for every \emph{sample point} in the line segment $\mathcal{L}(\bm x_{\alpha}, \bm x_{\beta})$, the line segment is recursively partitioned to generate new sample points, as Algorithm \ref{alg:addpoints} shows. 

\algrrule[0.8pt]
\begin{alg}
{Recursive partitioning of line segment $\mathcal{L}(\bm x_{\alpha}, \bm x_{\beta})$ in case that (\ref{cond_violate}) or (\ref{eq:violate2}) holds.}
\label{alg:addpoints}
\algrule[0.5pt]
\begin{algorithmic}[1]
\hspace{-4ex} \textbf{Input:} Linear MPC problem, initial  sample dataset $\mathcal{X}_1 \times \mathcal{U}_1$, the line segment $\mathcal{L}(\bm x_{\alpha}, \bm x_{\beta})$. \\
\hspace{-4ex} \textbf{Output:} Additional sample dataset $\mathcal{X}_2 \times \mathcal{U}_2$.
\STATE Initialize $flag=1$, $\mathcal{X}_2=\emptyset$, $\mathcal{U}_2=\emptyset$. 
\WHILE{flag}
\STATE $N_a=0$;
\FOR{$\bm x_i \in \mathcal{X}_1\cup \mathcal{X}_2$}
\STATE Select corresponding $u_i(\bm x_i) \in \mathcal{U}_1 \cup \mathcal{U}_2$ .
\IF {${\mathrm sign}(u_i(\bm x_i)-u_{i+1}(\bm x_i))=\mathrm{sign} (u_{i}(\bm x_{i+1})-u_{i+1}(\bm x_{i+1}))$}
\STATE $N_a=N_a+1$.
\STATE Add a point $\bm {x}_{\rm new} = 0.5(\bm x_i+\bm x_{i+1})$ to $\mathcal{X}_2$.
\STATE Calculate corresponding affine function $u^*(\bm x_{\rm new})$ through lines 7-9 of Algorithm 1, add to $\mathcal{U}_2$.
\ENDIF
\ENDFOR
\IF {Na=0}
\STATE flag=0.
\ENDIF
\ENDWHILE
\algrule[0.5pt]
\end{algorithmic}
\end{alg}

Lemma \ref{lem_add_points_line} shows that, if we add points according to Algorithm \ref{alg:addpoints},  (\ref{eq:assump_d}) and (\ref{eq:assump_c}) are satisfied for all the sample points in $\mathcal{L}(\bm x_{\alpha}, \bm x_{\beta})$.

\begin{lem}\label{lem_add_points_line}
Given the line segment $\mathcal{L}(\bm x_{\alpha}, \bm x_{\beta})$ such that (\ref{cond_violate}) or (\ref{eq:violate2}) holds, if we add points according to Algorithm \ref{alg:addpoints}, then we have
\begin{equation}\label{eq:conl_lem10}
\begin{array}{r}
\min\limits_{j \in J_{\geq,i}}u_j(\bm x_k)\leq u_k(\bm x_k),  \forall \bm x_i, \bm x_k \in \mathcal{L}(\bm x_{\alpha}, \bm x_{\beta}) \cap (\mathcal{X}_1\cup \mathcal{X}_2)
\end{array}
\end{equation}
and
\begin{equation}\label{eq:concl2_lem10}
\max\limits_{j \in J_{\leq,i}}u_j(\bm x_k)\geq u_k(\bm x_k), \forall \bm x_i, \bm x_k \in \mathcal{L}(\bm x_{\alpha}, \bm x_{\beta}) \cap (\mathcal{X}_1\cup \mathcal{X}_2).
\end{equation}
\end{lem}
\begin{pf}
After evaluating Algorithm \ref{alg:addpoints}, the condition
\begin{equation}\label{cond:uiuj}
(u_i(\bm x_i)-u_{i+1}(\bm x_i))\cdot (u_{i}(\bm x_{i+1})-u_{i+1}(\bm x_{i+1}))\leq 0
\end{equation}
is satisfied for all points $\bm x_i\in \mathcal{L}(\bm x_{\alpha}, \bm x_{\beta})$, which corresponds to two cases as shown in Fig. \ref{fig_uiuj}.  To generalize, here we only consider the case when $u_i \neq u_{i+1}$; for the case $u_i=u_{i+1}$, the affine function $u_i$ can connect the two points. Then, we can construct a continuous PWA function connecting the two points $\bm x_i$ and $\bm x_{i+1}$, i.e., $\min\{u_i, u_{i+1}\}$ for case 1 and $\max\{u_i, u_{i+1}\}$ for case 2. 
\begin{figure}[htbp]
\centering
 \psfrag{xi}[c]{$\bm x_i$}
 \psfrag{xj}[c]{$\bm x_{i+1}$}
  \psfrag{ui}[c]{$u_i$}
 \psfrag{uj}[c]{$u_{i+1}$}
   \subfigure[Case 1.]{
  \includegraphics[width=0.46\columnwidth]{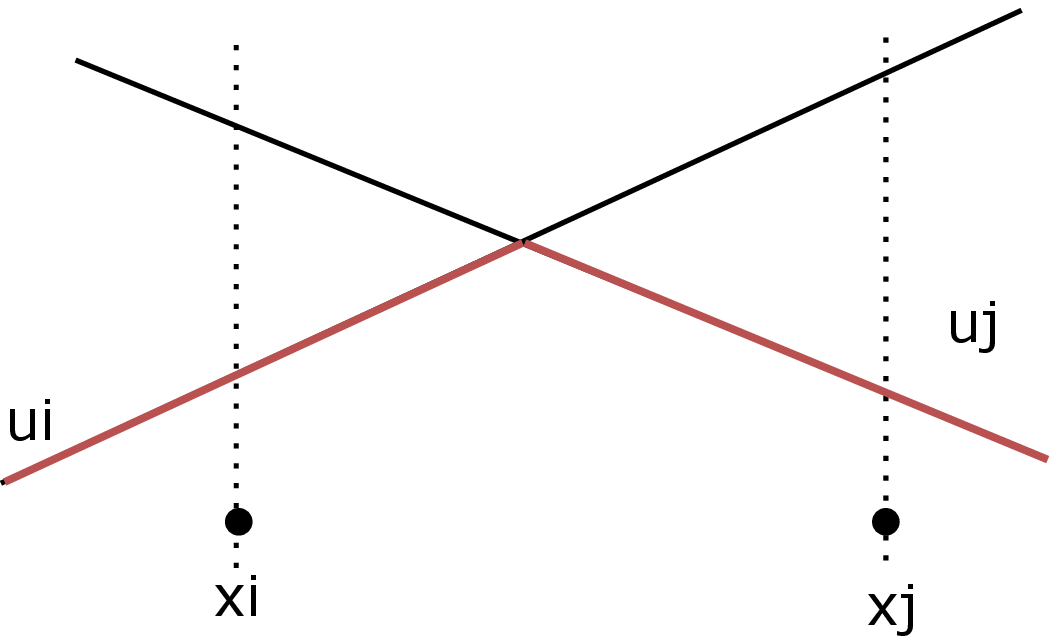}}
  \quad
  \subfigure[Case 2.]{
    \includegraphics[width=0.46\columnwidth]{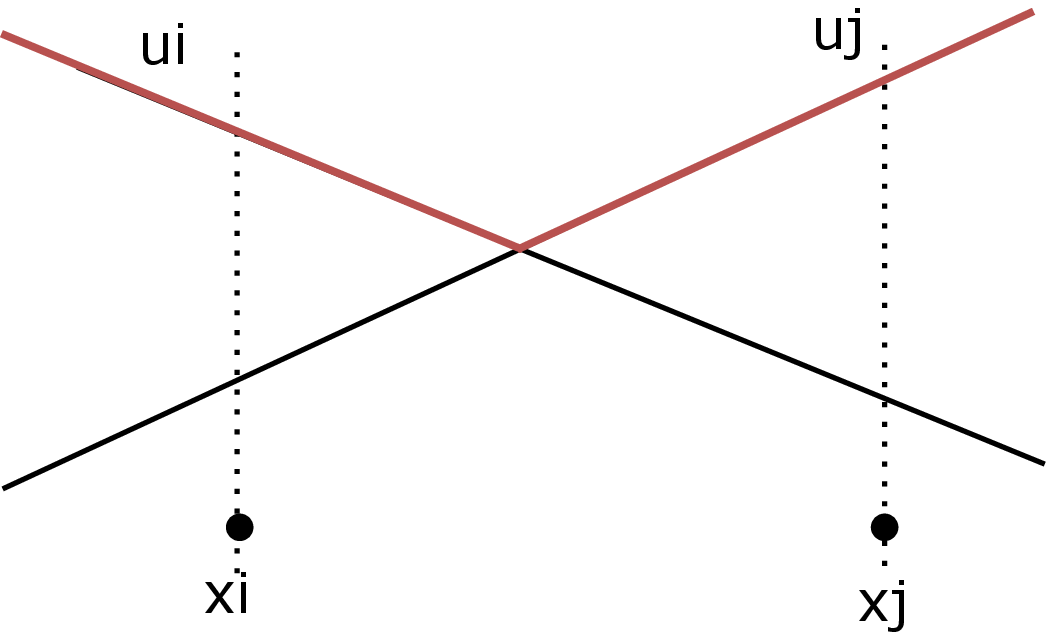}}
    \caption{Two cases satisfying (\ref{cond:uiuj}).}
  \label{fig_uiuj}
\end{figure}

Supposing that the constructed disjunctive and conjunctive continuous PWA functions connecting $\bm x_{\alpha}$ and $\bm x_{\beta}$ are $\hat{f}_{1}$ and $\hat{f}_2$, respectively, then we have 
\[
\hat{f}_1(\bm x_k)=\hat{f}_2(\bm x_k)=u_k(\bm x_k), \forall  \bm x_k \in \mathcal{L}(\bm x_{\alpha}, \bm x_{\beta}) \cap (\mathcal{X}_1\cup \mathcal{X}_2).
\]
Defining the index sets $\mathrm{aff}_1(\bm x_{\alpha}, \bm x_{\beta})$ and $\mathrm{aff}_1(\bm x_{\alpha}, \bm x_{\beta})$ as
 \[
 \mathrm{aff}_1{(\bm x_{\alpha}, \bm x_{\beta})}=\{j| \exists \bm x \in \mathcal{L}(\bm x_{\alpha}, \bm x_{\beta})~\mbox{such that}~\hat{f}_1(\bm x)=u_j(\bm x)\},
 \]
 and
  \[
 \mathrm{aff}_2{(\bm x_{\alpha}, \bm x_{\beta})}=\{j| \exists~ \bm x \in \mathcal{L}(\bm x_{\alpha}, \bm x_{\beta})~\mbox{such that}~\hat{f}_2(\bm x)=u_j(\bm x)\},
 \]
we then have
\[
\begin{array}{l}
\min\limits_{j \in J_{\geq,i}\cap \mathrm{aff}_1(\bm x_{\alpha}, \bm x_{\beta})}u_j(\bm x_k)\leq u_k(\bm x_k)\\
\quad\quad\quad\quad\quad\quad\quad\quad \forall \bm x_i, \bm x_k \in \mathcal{L}(\bm x_{\alpha}, \bm x_{\beta})\cap (\mathcal{X}_1\cup \mathcal{X}_2)
\end{array}
\]
and
\[
\begin{array}{l}
\max\limits_{j \in J_{\leq, i} \cap \mathrm{aff}_2(\bm x_{\alpha}, \bm x_{\beta})} u_j(\bm x_k) \geq u_k(\bm x_k)\\
\quad\quad\quad\quad\quad\quad\quad\quad\forall \bm x_i, \bm x_k \in \mathcal{L}(\bm x_{\alpha}, \bm x_{\beta})\cap (\mathcal{X}_1\cup \mathcal{X}_2).
\end{array}
\]

According to the above inequalities, we have
 (\ref{eq:conl_lem10}) and (\ref{eq:concl2_lem10}).

\end{pf}

Algorithm \ref{alg:addpoints} can be run repeatedly until, for all the sample points $(\bm x_i, u_i)\in \left(\mathcal{X}_1\cup \mathcal{X}_2\right) \times \left(\mathcal{U}_1\cup \mathcal{U}_2\right)$, we have (\ref{eq:assump_d}) and (\ref{eq:assump_c}); thus Assumption \ref{assump0} is satisfied and the resulting lattice PWA approximation equals the original control solution at all sample points and in the UO containing the sample points as interior points, as Lemma \ref{lem:lattice_approx} shows.

A simple one-dimensional example  is used to illustrate the process of constructing the disjunctive and conjunctive lattice PWA approximations and the re-sampling procedure.

\begin{exmp}\label{ex1}
Considering a one-dimensional continuous PWA function as in \cite{Xu2016irredundant},
\[
u(x)=\left\{\begin{array}{lr}
\ell_1(x)=0.5x+0.5&x \in [0, 1],\\
\ell_2(x)=2x-1&x \in [1, 1.5],\\
\ell_3(x)=2&x \in [1.5, 3.5],\\
\ell_4(x)=-2x+9&x \in [3.5, 4],\\
\ell_5(x)=-0.5x+3&x \in [4, 5],
\end{array}\right.
\]
the plot of which is shown in Fig. \ref{fig:ex1_plot}.
\begin{figure}[htbp]
\centering
\psfrag{x}[c]{$x$}
\psfrag{f(x)}[c]{$u(x)$}
\psfrag{l1}[c]{$\ell_1$}
\psfrag{l2}[c]{$\ell_2$}
\psfrag{l3}[c]{$\ell_3$}
\psfrag{l4}[c]{$\ell_4$}
\psfrag{l5}[l]{$\ell_5$}
\psfrag{x1}[c]{\tiny $(x_1, u(x_1))$}
\psfrag{x2}[c]{\tiny $(x_2, u(x_2))$}
\psfrag{x3}[c]{\tiny $(x_3, u(x_3))$}
\psfrag{x4}[c]{\tiny $(x_4, u(x_4))$}
\psfrag{x5}[c]{\tiny $(x_5, u(x_5))$}
\includegraphics[width=0.8\columnwidth]{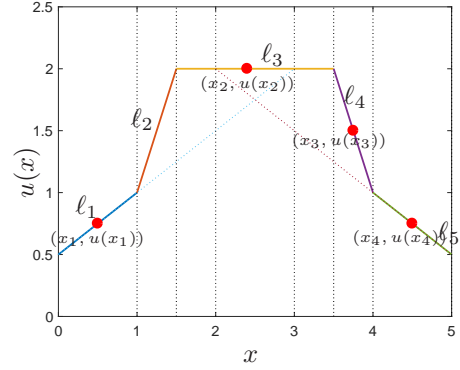}
\caption{One-dimensional continuous PWA function.}
\label{fig:ex1_plot}
\end{figure}

Supposing that we choose sample points as $(x_1, u(x_1))=(0.5, 0.75)$, $(x_2, u(x_2))=(2.5, 2)$, $(x_3, u(x_3))=(3.75, 1.5)$, and $(x_4, u(x_4))=(4.5, 0.75)$. Apparently, $(x_2, u(x_2))=(2.5, 2)$ is on the boundary of some UO region, as $\ell_1(2.5)=\ell_5(2.5)$, hence we apply a perturbation and move $x_2$ to 2.4, i.e., we have $(x_2, u(x_2))=(2.4, 2)$.

then the disjunctive lattice PWA approximation is
\[
\begin{array}{r}
\hat{f}_{\rm L, d}=\max\{\min\{\ell_1, \ell_3, \ell_4, \ell_5\}, \min\{\ell_3, \ell_4\},\\
 \min\{\ell_1, \ell_3, \ell_4\}, \min\{\ell_1, \ell_3, \ell_5\}\}.
 \end{array}
\]

The conjunctive lattice PWA approximation is constructed as
\[
\begin{array}{r}
\hat{f}_{\rm L, c}=\min\{\ell_1, \max\{\ell_3, \ell_5\}, \max\{\ell_4, \ell_5\}, 
\max\{\ell_4, \ell_5\}\}.
\end{array}
\]

It is apparent that the affine function $\ell_2$ has not been sampled, and we check whether (\ref{eq:assump_d}), (\ref{eq:assump_c}) are satisfied and find that
\[
\min\limits_{j \in J_{\geq, 2}}u_j(x_1)=\min\{\ell_3(x_1), \ell_4(x_1)\}>\ell_1(x_1),
\]
and
\[
\max\limits_{j \in J_{\leq, 1}}u_j(x_3)=\ell_1(x_3)<\ell_3(x_3).
\]
Hence more sample points should added between $x_1$ and $x_3$.

Applying Algorithm \ref{alg:addpoints}, we obtain the newly sampled points $x_5=1.2$ and $u(x_5)=2\times 1.2-1=1.4$ and now Assumption \ref{assump0} is satisfied.


Until now, all of the distinct affine functions $\ell_1, \ldots, \ell_5$ have been sampled, and we obtain the following lattice PWA approximations:
\[
\begin{array}{r}
\hat{f}_{\rm L, d}=\max\{\min\{\ell_1, \ell_3, \ell_4, \ell_5\}, \min\{\ell_2, \ell_3, \ell_4, \ell_5\},\\
\min\{\ell_2, \ell_3, \ell_4\}, \min\{\ell_1, \ell_2, \ell_3, \ell_4\}, \min\{\ell_1, \ell_2, \ell_3, \ell_5\}\}
 \end{array}
 \]
 and
 \[
 \begin{array}{r}
 \hat{f}_{\rm L, c}=\min\{\max\{\ell_1, \ell_2\}, \max\{\ell_1, \ell_2\},\\
 \max\{\ell_1, \ell_3, \ell_5\}, \max\{\ell_4, \ell_5\}, \max\{\ell_4, \ell_5\}\}.
 \end{array}
 \]
 For all $x$, as $\min\{\ell_2, \ell_3, \ell_4\}\geq \min\{\ell_2, \ell_3, \ell_4, \ell_5\}$ and $\min\{\ell_2, \ell_3, \ell_4\}\geq \min\{\ell_1, \ell_2, \ell_3, \ell_4\}$, the disjunctive approximation $\hat{f}_{\rm L, d}$ can be further expressed as
 \[
 \begin{array}{r}
 \hat{f}_{\rm L, d}=\max\{\min\{\ell_1, \ell_3, \ell_4, \ell_5\}, \min\{\ell_2, \ell_3, \ell_4\}, \\
 \min\{\ell_1, \ell_2, \ell_3, \ell_5\}\}.
 \end{array}
 \]
 Similarly, the conjunctive approximation can be rewritten as
 \[
 \hat{f}_{\rm L, c}=\min\{\max\{\ell_1, \ell_2\}, \max\{\ell_1, \ell_3, \ell_5\}, \max\{\ell_4, \ell_5\}\}.
 \]
 
 Fig. \ref{fig:ex1_fd} and \ref{fig:ex1_fc} give the plots of $\hat{f}_{\rm L, d}$ and $\hat{f}_{\rm L, c}$, respectively. It is apparent that for this simple example, after re-sampling, the disjunctive approximation $\hat{f}_{\rm L, d}$ equals $u(x)$, and there is \textbf{deviation} between the conjunctive approximation $\hat{f}_{\rm L, c}$ and $u(x)$. However, both $\hat{f}_{\rm L, d}$ and $\hat{f}_{\rm L, c}$ are identical to $u(x)$ at the sample points $x_1, \ldots, x_6$ and in $\Gamma(x_1), \ldots, \Gamma(x_6)$, confirming Lemma \ref{lem:lattice_approx}.
 \begin{figure}[htbp]
\centering
 \psfrag{x}[c]{$x$}
 \psfrag{fd}[c]{$\hat{f}_{\rm L, d}$}
\psfrag{x1}[c]{\tiny $(x_1, u(x_1))$}
\psfrag{x2}[c]{\tiny $(x_2, u(x_2))$}
\psfrag{x3}[c]{\tiny $(x_3, u(x_3))$}
\psfrag{x4}[c]{\tiny $(x_4, u(x_4))$}
\psfrag{x5}[c]{\tiny $(x_5, u(x_5))$}
  \includegraphics[width=0.8\columnwidth]{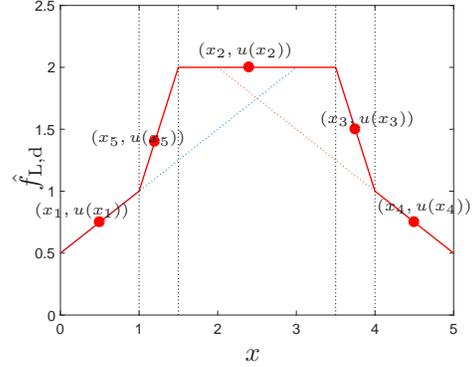}
     \caption{Disjunctive lattice PWA approximation.}
  \label{fig:ex1_fd}
\end{figure}
 \begin{figure}[htbp]
\centering
 \psfrag{x}[c]{$x$}
 \psfrag{fc}[c]{$\hat{f}_{\rm L, c}$}
\psfrag{x1}[c]{\tiny $(x_1, u(x_1))$}
\psfrag{x2}[c]{\tiny $(x_2, u(x_2))$}
\psfrag{x3}[c]{\tiny $(x_3, u(x_3))$}
\psfrag{x4}[c]{\tiny $(x_4, u(x_4))$}
\psfrag{x5}[c]{\tiny $(x_5, u(x_5))$}
  \includegraphics[width=0.8\columnwidth]{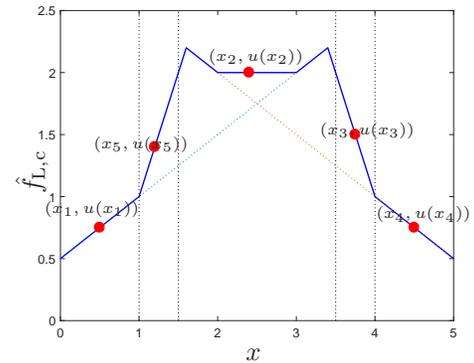}
     \caption{Conjunctive lattice PWA approximation.}
  \label{fig:ex1_fc}
\end{figure}

\end{exmp}

\subsection{Simplification of lattice PWA approximation}
In Example \ref{ex1}, duplicated or redundant terms have been removed from the lattice PWA approximation, which simplifies the approximation. When the number of sample points $N_s$ is large, the evaluation of (\ref{eq:lattice_approximation_d}) and (\ref{eq:lattice_approximation_c}) are not easy, and hence the simplification is considered in this  section. 

The simplification of a disjunctive lattice PWA function was addressed in \cite{Xu2016irredundant}, for which the detailed subregions of the PWA function are known. In this paper, the information of the subregions of the PWA function is unknown. It is also difficult to derive the expression of the subregion polyhedra through lattice PWA approximation. 
\cbstart
Hence, in this section, the disjunctive and conjunctive lattice PWA approximations are simplified according to the following rule. Assuming that the set $C$ denotes the codomain of affine functions $u_1, u_2, \ldots$, the operations $\bigvee$ and $\bigwedge$ are defined as follows,
\[
u_i \bigvee u_j = \max\{u_i, u_j\}, u_i \bigwedge u_j = \min\{u_i, u_j\}.
\]
It has been shown in  \cite{Tarela1975minimization} that 
the set $C$, together with the operations $\bigvee$ and $\bigwedge$, constitutes a distributive lattice, and the following property holds for all $u_i, u_j \in C$:
\begin{equation}\label{eq:axiomR1}
R1: \begin{array}{l}
u_i \bigvee (u_i \bigwedge u_j)=u_i\\
u_i \bigwedge (u_i \bigvee u_j)=u_i.
\end{array}
\end{equation}
Actually, the above rule has been used in the simplification of lattice PWA approximations $\hat{f}_{\rm L, d}$ and $\hat{f}_{\rm L, c}$ in Example \ref{ex1}. After simplification, we get lattice PWA approximations with far less terms, which are equivalent to the lattice PWA approximations constructed through (\ref{eq:lattice_approximation_d}) and (\ref{eq:lattice_approximation_c}).
\cbend

The process of obtaining disjunctive and conjunctive lattice PWA approximations and simplifying them can be summarized in Algorithm \ref{alg:total}.

\algrrule[0.8pt]
\begin{alg}
{Construction and simplification of disjunctive and conjunctive lattice PWA approximations.}
\label{alg:total}
\algrule[0.5pt]
\begin{algorithmic}[1]
\hspace{-4ex} \textbf{Input:} Linear MPC problem. \\
\hspace{-4ex} \textbf{Output:} Simplified disjunctive and conjunctive lattice PWA approximations.
\STATE Generate sample dataset $\mathcal{X}_1\times \mathcal{U}_1$ according to Algorithm \ref{alg:sampling}. 
\STATE Generate additional sample dataset $\mathcal{X}_2 \times \mathcal{U}_2$ according to Algorithm \ref{alg:addpoints}.
\FOR{$\bm x_i \in \mathcal{X}_1\cup \mathcal{X}_2$}
\STATE Calculate the index sets $J_{\geq, i}$ and $J_{\leq, i}$ at the point $\bm x_i$ according to Equation (\ref{eq:index_Jgeq}) and (\ref{eq:index_Jleq}), respectively.
\ENDFOR
\STATE Construct the disjunctive and conjunctive lattice PWA approximations according to Equation (\ref{eq:lattice_approximation_d}) and (\ref{eq:lattice_approximation_c}).
\STATE Simplify the lattice PWA approximations according to the rule R1 (\ref{eq:axiomR1}).
\algrule[0.5pt]
\end{algorithmic}
\end{alg}

The  disjunctive  and conjunctive lattice PWA approximations of explicit MPC control law are illustrated using a small example of a linear discrete-time system. 
\begin{exmp}\label{ex2}
Considering the discrete-time double integrator example introduced in \cite{Johansson2003piecewise}, the system dynamics can be written as
\[
\begin{array}{rcl}
x_{k+1}&=&\left[\begin{array}{cc}
1&T_s\\
0&1
\end{array}\right]+\left[\begin{array}{c}
T_s^2\\
T_s
\end{array}\right]u_k
\end{array}
\]
where the sampling interval $T_s$ is 0.3. Considering the MPC problem with $Q=\mathrm{diag}(1,0)$, $R=1$, and $P$ is the solution of the discrete-time algebraic Riccati equation. The  system constraints are $-1 \leq u_k \leq 1$ and $-0.5 \leq x_{k, 2}\leq 0.5$. The region is set to be $[-2.8, 2.8] \times [-0.8, 0.8]$.

To derive the lattice PWA approximation, 441 points ($21\times 21$) are uniformly generated in the region $[-1,1]^2$. For all 441 points, 68 points reach the boundary of some subregions, and we apply perturbations to move the points to the interior of some UO regions. 
There are five distinct affine functions, which are
\[
\begin{array}{l}
u_1=-0.8082x_1-1.1559x_2;\\
u_2 = -3.3333x_2-2.6667;\\
u_3=-3.3333x_2+2.6667; \\
u_4 = -1;  u_5=1.
\end{array}
\] 
We generate 441 terms, and after the simplification of terms, the following lattice PWA approximations are obtained:
\begin{equation*}
\hat{f}_{\mathrm{L, d}}(\bm x)=\max\{u_2(\bm x), u_4(\bm x), \min(u_1(\bm x), u_3(\bm x), u_5(\bm x))\}
\end{equation*}
and
\[
\hat{f}_{\mathrm{L, c}}(\bm x)=\min\{\max(u_1(\bm x), u_2(\bm x), u_4(\bm x)), u_3(\bm x), u_5(\bm x)\}.
\]
Readers can verify that the disjunctive and conjunctive approximations are equivalent in $[-2.8, 2.8] \times [-0.8, 0.8]$. Fig. \ref{fig_ex2}(a) gives the optimal MPC controller generated by the MPT3 Toolbox \cite{MPT3}, and the linear subregions are also shown, see Fig. \ref{fig_ex2}(b). For this example, the lattice PWA approximations are identical to explicit MPC. In the next section, we demonstrate that if all the affine functions have been sampled and the two lattice PWA approximations are identical, then both of them equal the optimal MPC control law.
\begin{figure}[htbp]
\centering
 \psfrag{x1}[c]{$x_1$}
 \psfrag{x2}[c]{$x_2$}
 \psfrag{u}[c]{$u^*$}
   \subfigure[Controller.]{
  \includegraphics[width=0.47\columnwidth]{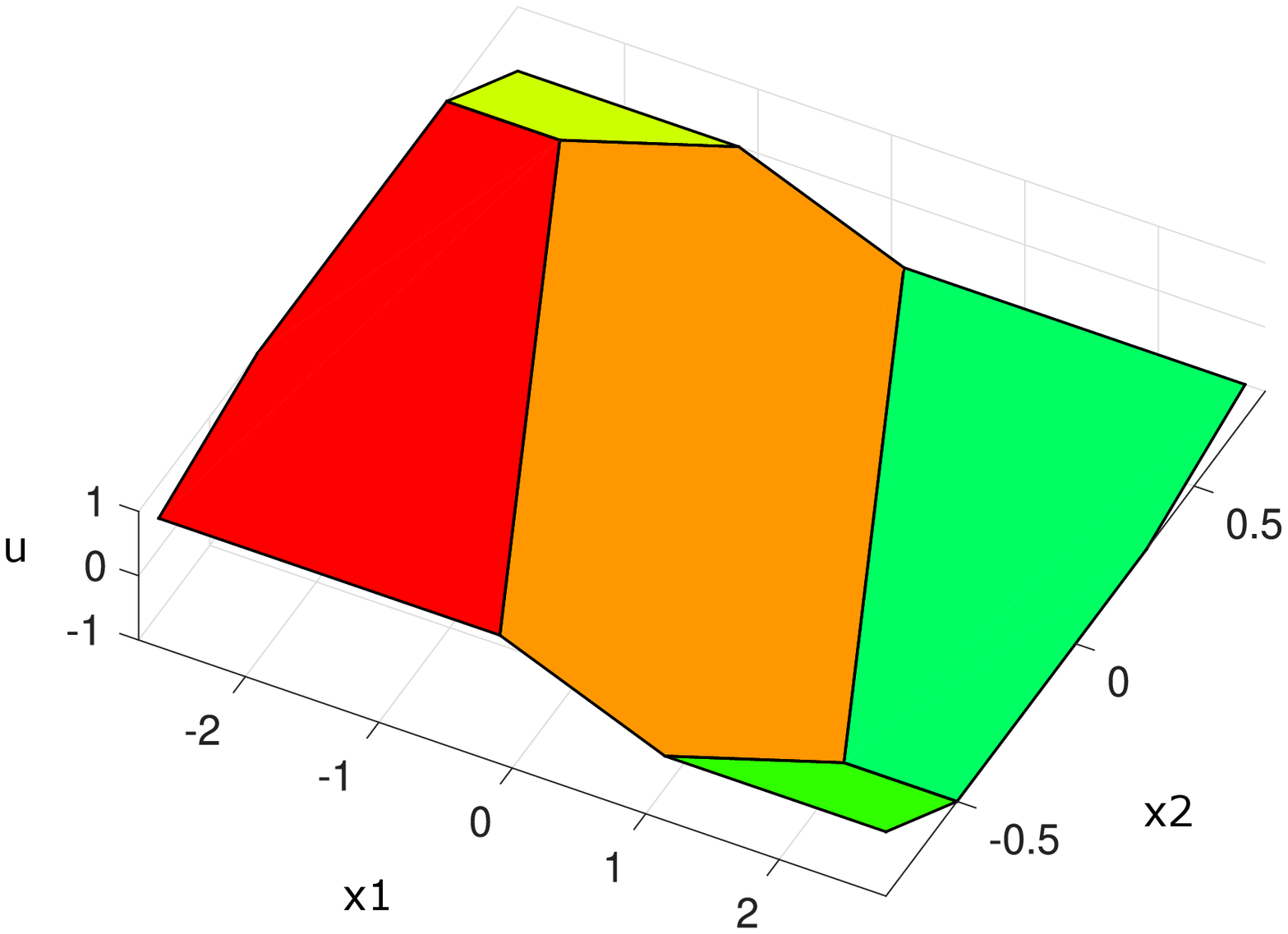}}
  \subfigure[Region.]{
    \includegraphics[width=0.47\columnwidth]{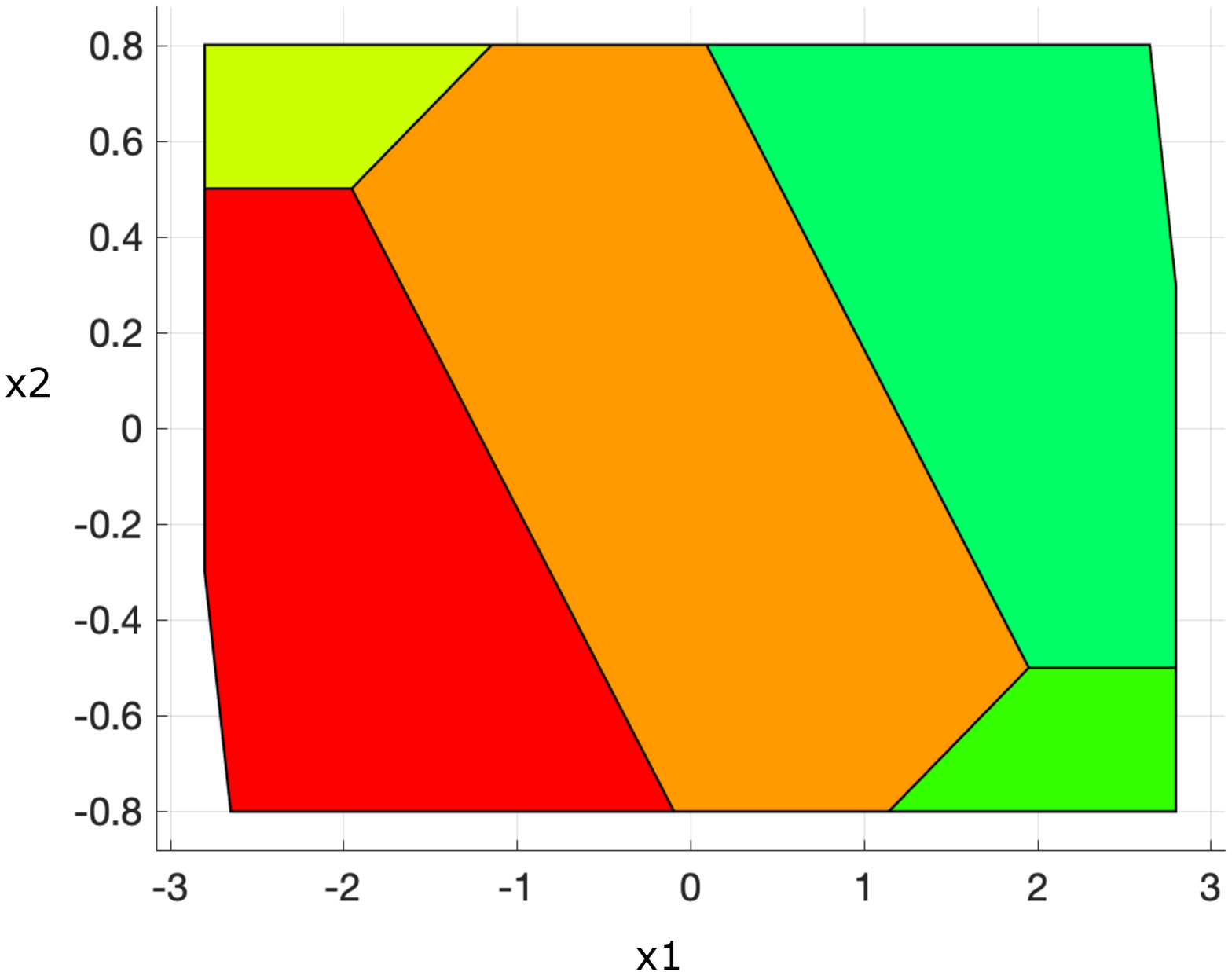}
    }
    \caption{Explicit MPC controller in Example \ref{ex2}.}
  \label{fig_ex2}
\end{figure}
\end{exmp}

\section{Approximation error and computational complexity}

\subsection{Deviations between the disjunctive and conjunctive approximations}

After evaluating Algorithm \ref{alg:sampling}-\ref{alg:addpoints}, we obtain the sample point set $(\mathcal{X}_1\cup \mathcal{X}_2
)\times (\mathcal{U}_1 \cup \mathcal{U}_2 )$, and use $\mathcal{X}=\mathcal{X}_1\cup \mathcal{X}_2
$ and $\mathcal{U}=\mathcal{U}_1 \cup \mathcal{U}_2 $ to denote the sample point set. Assuming that $\mathcal{X}=\{\bm x_1, \ldots, \bm x_{N_s}\}$, given both the disjunctive and conjunctive lattice PWA approximations, under the following assumption, the deviation between the approximations and the optimal control law can be derived.
\cbstart
\begin{assum}\label{assump1}
We assume that all the distinct affine functions have been sampled.
\end{assum}

All the distinct affine functions for a PWA MPC optimal controller can be obtained by collecting all critical regions and corresponding local affine functions $u_{\rm loc}(x)$ as defined in Definition 1 and selecting the distinct ones. 
\cbend

Assumption \ref{assump1} can be explained as follows. Suppose we have 3 regions $\Omega_1, \Omega_2, \Omega_3$ and the local affine functions $u_{\rm loc, 1}(\bm x)=\ell_1(\bm x), u_{\rm loc, 2}(\bm x)=\ell_1(\bm x), u_{\rm loc, 3}(\bm x)=\ell_2(\bm x)$, then the distinct affine functions are $\ell_1(\bm x)$ and $\ell_2(\bm x)$, and by sampling points in $\Omega_1$ and $\Omega_3$, we have sampled all the distinct affine functions.

Theorem \ref{thm:1} bounds the error between the lattice PWA approximations and the original optimal control law.
\begin{thm}\label{thm:1}
Supposing that
\[
\hat{f}_{\rm L, d}(\bm x)=\max\limits_{i \in \{1, \ldots, N_s\}}\min\limits_{j \in J_{\geq, i}}u_j(\bm x)
\] and 
\[
\hat{f}_{\rm L, c}(\bm x)=\min\limits_{i \in \{1, \ldots, N_s\}}\max\limits_{i \in J_{\leq, i}}u_i(\bm x)
\] are the disjunctive and conjunctive approximations of the optimal control law $u^*(\bm x)$ over the domain $\Omega$, assuming that Assumption \ref{assump1} holds, and defining
\begin{equation}\label{eq:epsilon}
\varepsilon = \max\limits_{\bm x}\left(\hat{f}_{\rm L, c}(\bm x)-\hat{f}_{\rm L, d}(\bm x)\right),
\end{equation}
we then have
\begin{equation}\label{eq:concl1_thm2}
-\varepsilon\leq \hat{f}_{\rm L, d}(\bm x)-u^*(\bm x) \leq 0
\end{equation}
and
\begin{equation}\label{eq:concl2_thm2}
0\leq \hat{f}_{\rm L, c}(\bm x)-u^*(\bm x) \leq \varepsilon.
\end{equation}
Furthermore, if $\varepsilon=0$, we have
\begin{equation}\label{eq:identical}
\hat{f}_{\rm L, d}(\bm x)=\hat{f}_{\rm L, c}(\bm x)=u^*(x), \forall \bm x \in \Omega.
\end{equation}
\end{thm}
\begin{pf}
Assume $\mathcal{N}$ is the index of all UO regions, according to the conclusion in \cite{Xu2016irredundant}, we have
\begin{equation}\label{eq:full_lattice_d}
u^*(\bm x)=\max\limits_{i \in \mathcal{N}}\min\limits_{j \in I_{\geq,i}}u_j(\bm x), \forall \bm x \in \Omega,
\end{equation}
\begin{equation}\label{eq:full_lattice_c}
u^*(\bm x)=\min\limits_{i \in \mathcal{N}}\max\limits_{j \in I_{\leq, i}}u_j(\bm x), \forall \bm x \in \Omega.
\end{equation}
in which the index sets $I_{\geq, i}$ and $I_{\leq, i}$ are defined as,
\[
I_{\geq,i}=\{j|u_j(\bm x)\geq u_i(\bm x), \forall \bm x \in \Gamma(\bm x_i)\},
\]
\[
I_{\leq, i} =\{j|u_j(\bm x)\leq u_i(\bm x), \forall \bm x \in \Gamma(\bm x_i)\}.
\]

If Assumption \ref{assump1} holds, i.e., all the distinct affine functions have been sampled, then for a UO region $\Gamma(\bm x_i)$,  as the order of affine functions remains unchanged in the UO region, the set $I_{\geq, i}$
is identical to $J_{\geq,i}$ defined in (\ref{eq:index_Jgeq}). Similarly, the sets $I_{\leq, i}$ and $J_{\leq, i}$ defined in (\ref{eq:index_Jleq}) are equivalent. 
Therefore, for all $\bm x \in \Omega$ and all $i \in \mathcal{N}$, the following inequalities hold:
\begin{equation}\label{eq:term_less_f}
\min\limits_{j \in J_{\geq,i}}u_j(\bm x)\leq u^*(\bm x),
\end{equation}
and
\begin{equation}\label{eq:term_ge_f}
\max\limits_{j \in J_{\leq, i}} u_j(\bm x) \geq u^*(\bm x).
\end{equation}

As the sampled UO regions are only a subset of all UO regions, i.e.,
\[
\{1, \ldots, N_s\} \subset \mathcal{N},
\]
we have inequalities (\ref{eq:term_less_f}) and (\ref{eq:term_ge_f})  for all $\bm x \in \Omega$ and all $i \in \{1, \ldots, N_s\}$.


Then the following is valid:
\[
\hat{f}_{\rm L, d}(\bm x) \leq u^*(\bm x)\leq \hat{f}_{\rm L, c}(\bm x) , \forall \bm x\in \Omega.
\]
 According to (\ref{eq:epsilon}), we have (\ref{eq:concl1_thm2}) and (\ref{eq:concl2_thm2}).

Furthermore, if $\varepsilon=0$, then both approximations are identical to the optimal control law in the region $\Omega$, i.e., (\ref{eq:identical}) holds.
\end{pf}

As Theorem \ref{thm:1} indicates, in order to guarantee that the lattice PWA approximations are error-free, two conditions should be satisfied, the first is the validity of Assumption \ref{assump1}, and the second is that $\varepsilon$ as defined in (\ref{eq:epsilon}) is zero.
 
\cbstart
\begin{rem}
As a necessary condition for the validity of (\ref{eq:identical}), i.e., the lattice PWA approximations are error-free, Assumption \ref{assump1} requires that all the distinct affine functions have been sampled. It is noted that this is not the same as the condition that all the critical regions should be identified. As \cite{Xu2016irredundant} shows, the number of distinct affine functions in explicit MPC is generally far more less than the number of critical regions, as some critical regions can share the same affine function.
\end{rem}
\cbend

\subsubsection{A necessary condition for the validity of Assumption \ref{assump1}}
In general, it is not easy to check whether Assumption \ref{assump1} is satisfied, which is vital for the validity of (\ref{eq:term_less_f}) and (\ref{eq:term_ge_f}) for all $\bm x \in \Omega$ and $i \in \{1, \ldots, N_s\}$. Hence we check whether the relaxed condition
\begin{equation}\label{eq:cond_gen}
\min_{j \in J_{\geq, i}}u_j(\bm x)\leq \max_{j \in J_{\leq, i}}u_j(\bm x), \forall i \in \{1, \ldots, N_s\}, \forall \bm x \in \Omega
\end{equation}
holds, which is a direct result of (\ref{eq:term_less_f}) and (\ref{eq:term_ge_f}).

To check whether (\ref{eq:cond_gen}) is satisfied, for any $i,k \in \{1, \ldots, N_s\}$, the following optimization problem is solved:
\begin{equation}\label{eq:optimizaiton40}
\begin{array}{rl}
\min\limits_{\bm x} & \max\limits_{j \in J_{\leq, i}}u_j(\bm x)-\min\limits_{j \in J_{\geq, k}}u_j(\bm x)\\
s.t. & \bm x \in \Omega.
\end{array}
\end{equation}
 If the optimal value for all $i,k \in \{1, \ldots, N_s\}$ is nonnegative, then (\ref{eq:cond_gen}) holds. 

The optimization problem (\ref{eq:optimizaiton40}) can be transformed into an equivalent linear programming  (LP) problem:
\begin{equation}\label{eq:lp}
\begin{array}{rl}
\min\limits_{\bm x, y_1, y_2} &y_1+y_2 \\
s.t. & \bm x \in \Omega,\\
{}&u_j(\bm x)\leq y_1, \forall j \in J_{\leq, i}, \forall i \in \{1, \ldots, N_s\}\\
{}&u_j(\bm x)\geq -y_2, \forall j \in J_{\geq, k}, \forall k \in \{1, \ldots, N_s\}
\end{array}
\end{equation}
which is easy to solve as $\Omega$ is a polyhedron.
If we find a point $\bm x$ and an index pair $i, k$, such that the cost in (\ref{eq:lp}) is negative, which means that (\ref{eq:cond_gen}) is violated, we can generate more sample points and get more distinct affine functions, as Lemma \ref{lem16} shows.

\begin{lem}\label{lem16}
Suppose there are some $\bm x_{\gamma}$ and indices $\alpha, \beta$ such that $y_1(\bm x_{\gamma})+y_2(\bm x_{\gamma})<0$, then more sample points can be generated  in the line segment $\mathcal{L}(\bm x_i, \bm x_{\gamma})$ or $\mathcal{L}(\bm x_{\gamma}, \bm x_k)$ according to Algorithm \ref{alg:addpoints}, and more distinct affine functions can be sampled.
\end{lem}
\begin{pf}
As $y_1(\bm x_{\gamma})+y_2(\bm x_{\gamma})<0$ for some $\alpha, \beta$, we have 
\begin{equation}\label{eq:dlargerc}
\max\limits_{j \in J_{\leq, \alpha}}u_j(\bm x_{\gamma})<\min\limits_{j \in J_{\geq, \beta}}u_k(\bm x_{\gamma}).
\end{equation}

Then at least one of the inequalities 
\begin{equation}\label{eq:dlargeru}
\min\limits_{j \in J_{\geq, \alpha}}u_j(\bm x_{\gamma})>u^*(\bm x_{\gamma})
\end{equation}
and
\begin{equation}\label{eq:clessu}
\max\limits_{j \in J_{\leq, \beta}}u_j(\bm x_{\gamma})<u^*(\bm x_{\gamma})
\end{equation}
is valid.
This is apparent, since, if both (\ref{eq:dlargeru}) and (\ref{eq:clessu}) do not hold, we have 
\[
\min\limits_{j \in J_{\geq, \alpha}}u_j(\bm x_{\gamma})\leq \max\limits_{j \in J_{\leq, \beta}}u_j(\bm x_{\gamma}),
\]
which contradicts (\ref{eq:dlargerc}).

\cbstart
The checking of (\ref{eq:dlargeru}) and and (\ref{eq:clessu}) is very simple, as $\bm x_{\gamma}$ and $u^*(\bm x_{\gamma})$ are both known, and computing $\min\limits_{j \in J_{\geq, \alpha}}u_j(\bm x_{\gamma})$ only needs arithmetic operations, i.e., comparing the functions values of several affine functions at the point $\bm x_\gamma$. \cbend 
If (\ref{eq:dlargeru}) holds, then according to Lemma \ref{lem_add_points_line}, sample points can be added to the line segment $\mathcal{L}(\bm x_{\alpha}, \bm x_{\gamma})$ as in Section \ref{sec:guarantee1} to ensure that
\[
\min\limits_{j \in J_{\geq,i}}u_j(\bm x_k)\leq u_k(\bm x_k)
\]
for all the sample points in $\mathcal{L}(\bm x_{\alpha}, \bm x_{\gamma})$. In this process, more distinct affine functions have been identified.

For the conjunctive case, if (\ref{eq:clessu}) is valid, we can also add sample points in the line segment $\mathcal{L}(\bm x_{\beta}, \bm x_{\gamma})$ according to Lemma \ref{lem_add_points_line} such that
\[
\max\limits_{j \in J_{\leq,i}}u_j(\bm x_k)\geq u_k(\bm x_k)
\]
for all the sample points in $\mathcal{L}(\bm x_{\beta}, \bm x_{\gamma})$.

By doing so, more sample points as well as more distinct affine functions have been sampled, which is necessary for the validity of Assumption \ref{assump1}.
\end{pf}

\begin{rem}
 It is noted that  although the relaxed condition (\ref{eq:cond_gen}) is only a necessary condition for the validity of Assumption \ref{assump1}, the checking of (\ref{eq:cond_gen}) is with respect to the entire domain $\Omega$, not only the sample points. And we discovered in our numerical experiments that in most cases, if (\ref{eq:cond_gen}) holds in the domain of interest, all the distinct affine functions active in that region have been identified.
\end{rem}

\subsubsection{Checking whether the 2 approximations are identical}

Both the disjunctive approximation $\hat{f}_{\rm L, d}$ and conjunctive approximation $\hat{f}_{\rm L, c}$ are continuous PWA functions, so whether $\hat{f}_{\rm L, d}=\hat{f}_{\rm L, c}$ can be checked in each linear subregion of both $\hat{f}_{\rm L, d}$ and $\hat{f}_{\rm L, C}$. When the number of literals and terms in the lattice PWA approximations are large, it is not easy to identify all the linear subregions. So here we resort to a statistical method, i.e., generating a huge number of i.i.d. sample points, which constitute a validation dataset $\mathcal{X}_{\rm validate}=\{\bm x_i, {i=1}, \ldots, {N_v}\}$. For each sample point, we define an indicator function as
\[
I(\bm x_i):=\left\{
\begin{array}{cc}
1&\mbox{if}~ \hat{f}_{\rm L, d}(\bm x_i)=\hat{f}_{\rm L, c}(\bm x_i)\\
0& \mbox{if}~ \hat{f}_{\rm L, d}(\bm x_i)\neq \hat{f}_{\rm L, c}(\bm x_i),
\end{array}
\right.
\]
and then the random variables $I(\bm x_i), i=1,\ldots, N_v$ are also i.i.d.  Denoting the probability for $I(\bm x_i=1)$ as $\mu$, i.e., $\mathbb{P}[I(\bm x_i)=1]=\mu$, then according to Hoeffding's inequality \cite{Hoeffding1994probability}, we have
\[
\mathbb{P}[|\mu-\bar{I}|\geq \epsilon]\leq 2\exp(-2N_v \epsilon^2),
\]
in which $\bar{I}=\frac{1}{N_v}\sum\limits_{k=1}^{N_v}I(\bm x_k)$. Therefore, we have
\[
\mathbb{P}[\mu \geq \bar{I}-\epsilon]>1-2\exp(-2N_v \epsilon^2),
\]
meaning that with confidence  $1-2\exp(-2N_v \epsilon^2)$, the probability that the lattice PWA approximations $\hat{f}_{\rm L, d}$ and $\hat{f}_{\rm L, c}$ are identical is larger than $\bar{I}-\epsilon$. If $\bar{I}=1$, then
by setting a small enough threshold $\epsilon$, we can say that   with confidence $1- 2\exp(-2N_v \epsilon^2)$, the lattice PWA approximations $\hat{f}_{\rm L, d}$ and $\hat{f}_{\rm L, c}$ are almost identical, and thus both equal the optimal control law. For example, if $\epsilon=10^{-3}$, then $N_v\geq 5\times 10^6$ can ensure that the confidence is almost 1.

\subsection{Complexity analysis}

\subsubsection{Online evaluation}\label{sec:online}

Assuming that there are $\tilde{N}$ terms in the final approximation, according to \cite{Wen2009analytical}, the worst-case online evaluation complexity is $O(\tilde{N}^2)$. In general, we have $\tilde{N}\ll N_s$, and hence the online evaluation is very fast.

\subsubsection{Storage requirements}

Assuming that the disjunctive lattice PWA approximation has $\tilde{N}$ terms and $M$ literals (i.e. the number of distinct affine functions is $M$), we must store $(n_x+1)\cdot M$ real numbers and $\sum\limits_{i=1}^{\tilde{N}}|J_{\geq, i}|$ integer numbers, in which $|J_{\geq, i}|$ is the number of elements in the set $J_{\geq, i}$. 
As $|J_{\geq,i} |\leq M$,  in total  $(n_x+1)\cdot M$ real numbers and $M\cdot \tilde{N}$ integer numbers must be stored.

In many  cases, we have $\tilde{N}\ll N_s$, and hence the storage requirement for the  disjunctive lattice PWA approximation is very small. 

For the conjunctive lattice PWA approximation, we achieve the same result.

\subsubsection{Offline complexity}

The offline time complexity for deriving equivalent disjunctive and conjunctive lattice PWA approximations can be summarized as follows.

The time complexity consists of two parts. One concerns the training points sampling and re-sampling in order to obtain lattice PWA approximations and the  other is the complexity of checking whether the lattice PWA approximations are error-free. Lemma \ref{lem:last} describes the worst-case offline time complexity.

\begin{lem}\label{lem:last}
Assuming that the sample domain is a hyperbox, then the worst-case complexity of deriving the disjunctive and conjunctive lattice PWA approximations is $O(N_v\cdot N_s^2)$, in which $N_s$ and $N_v$ are the numbers of sample points in $\mathcal{X}$ and $\mathcal{X}_{\rm validate}$, respectively.
\begin{pf}
As indicated previously, the offline complexity comes from evaluating Algorithm \ref{alg:total} and checking whether the two lattice PWA approximations are equivalent. 

For Algorithm \ref{alg:total}, the evaluation of Algorithm \ref{alg:sampling} and \ref{alg:addpoints}, and the simplification of lattice PWA approximations account for the major part of the computational complexity. 

The number of sample points in Algorithm \ref{alg:sampling} can be calculated as $N_1=\prod\limits_{i=1}^n \frac{b_i-a_i}{\delta_i}$, in which $[a_i, b_i]$ is the sample range for the $i$-th component, and correspondingly, $\delta_i$ is the length of the sample interval. The complexity of evaluating Algorithm \ref{alg:sampling} includes solving $N_1$ convex quadratic programming problems and corresponding KKT conditions, which are basically solving linear equations. 
The time complexity of solving $N_1$ convex quadratic programming problems with $N_p\cdot n_u$ decision variables is approximately $O(N_1 \cdot L^2(N_p\cdot n_u)^4)$ by using an interior-point algorithm \cite{Ye1989extension}, in which $L$ is the bit length of the quadratic programming problem. The dominant algorithmic operation in solving the KKT conditions is solving $N_1$ matrix inversion problems, the worst-case complexity of which is $O(N_1|\mathcal{A}^*|^3)$ using the Gauss-Jordan elimination algorithm, in which $|\mathcal{A}^*|$ is the number of active constraints. As $|\mathcal{A}^*|\leq p$, where $p$ is the number of constraints in QP (\ref{mp-qp2}), the worst-case complexity for solving the KKT conditions is $O(N_1 p^3)$.   

We now discuss the worst-case complexity of evaluating Algorithm \ref{alg:addpoints}. For two points $\bm x_{\alpha}$ and $\bm x_{\beta}$, if (\ref{eq:term_less_f}) is violated, the evaluation of Algorithm  \ref{alg:addpoints} is basically a binary search method for identifying omitted subregions in the line segment $\mathcal{L}(\bm x_{\alpha}, \bm x_{\beta})$ as well as calculating affine functions for newly generated points. Assume the maximum number of subregions  appearing in $\mathcal{L}(\bm x_{\alpha}, \bm x_{\beta})$ is $M_{\alpha, \beta}$, and generally $M_{\alpha, \beta}<N_1$.
The binary searching of the subregions then yields a worst-case complexity of $O(\log_2N_1)$ and the affine function calculation yields a worst-case complexity of $O(p^3\log_2N_1)$, in which $p$ is the number of constraints in QP (\ref{mp-qp2}). Supposing that there are $N_t$ point pairs such that (\ref{eq:term_less_f}) is violated, then the worst-case complexity is  $O(N_t \cdot p^3\log_2 N_1)$. In general, the number $N_t$ is closely related to the sample grid size $\delta_i$, which then depends on the number of sample points $N_1$ in Algorithm 1. A larger $N_1$ will result in a smaller number of $N_t$; hence, the complexity of Algorithm \ref{alg:addpoints} can be decreased by increasing the complexity of Algorithm 1.  



After evaluating Algorithm \ref{alg:sampling}-\ref{alg:addpoints}, there are $N_s$ sample points. The simplification procedure requires the comparison of the sets $J_{\geq, i}$ ($J_{\leq, i}$) for $i=1,  \ldots, N_s$, which at most yields $\left(^{~2}_{N_s}\right)=\frac{N_s(N_s-1)}{2}$ times comparisons. For each comparison, at most $M^2$ literals need to be compared. Hence the worst-case complexity for the simplification is $O(M^2N_s^2)$. 

After simplification, we have only $\tilde{N}$ terms, which is generally much less than $N_s$, i.e., $\tilde{N}\ll N_s$, then for the process of checking whether the lattice PWA approximations are error-free, one has to solve at most $\tilde{N}$ LP problems and generate $N_v$ validation points. Assuming that $L$ is the bit length of the LP problem (\ref{eq:lp}), then the worst-case complexity for solving $\tilde{N}^2$ LP problems is $O(\tilde{N}^2 \cdot n_x^{3.5}\cdot L)$. For the sampling of  $N_v$ points in the lattice PWA approximations, as indicated in Section \ref{sec:online}, the function evaluation process of lattice PWA approximations has a worst-case complexity of $O(\tilde{N}^2)$, in which $\tilde{N}<N_s$, hence the worst-case complexity of generating validation points is $O(N_v\cdot \tilde{N}^2)$.
As $N_v$ is large, we have $O(\tilde{N}^2 \cdot n_x^{3.5}\cdot L)\ll O(N_v\cdot \tilde{N}^2)$; hence the worst-case complexity of validation is $O(N_v\cdot \tilde{N}^2)$.

In general, $N_1<N_s$, $p\ll N_1$, $N_t \ll N_1$, $M\ll N_v$, and $\tilde{N}\ll N_s$, and then the  total worst-case complexity is $O(N_v \cdot N_s^2)$.

\end{pf}

\end{lem}

It is noted that $O(N_v \cdot N_s^2)$ is the worst-case complexity. In the simulation results, we can see that the offline calculation is actually not time consuming.

\section{Simulation results}
\begin{exmp}\label{example2}
Consider an example taken from \cite{Karg2020efficient}, which is the inverted pendulum on a cart. The state consists of the angle and the angle speed of the pole, i.e., $\Phi$, $\dot{\Phi}$, respectively, and the position and speed  of the cart, i.e., $s$ and $\dot{s}$, respectively.  The constraints for the state are $\|\bm x\|_{\infty}^T \leq [1, 1.5, 0.35, 1.0]^T$. The input is the force, the constraint of which is $|u|\leq 1$. The discrete-time dynamic is given by
\[
A=\left[
\begin{array}{cccc}
1&0.1&0&0\\
0&0.9818&0.2673&0\\
0&0&1&0.1\\
0&-0.0455&3.1182&1
\end{array}
\right], B=\left[\begin{array}{c}
0\\
0.1818\\
0\\
0.4546
\end{array}\right].
\]
The prediction horizon is taken to be $N=10$. The value of matrices in the cost function is $Q = \rm diag\{2, 2, 2, 2\}$, $R=1$, and $P = 0$. According to the MPT3 toolbox, the optimal control solution is a PWA function of the state $\bm x$, with 2,271 polyhedral regions.

In \cite{Karg2020efficient}, 88,341 samples were generated to train an approximated controller, which is basically a deep PWL neural network. Here, 
in order to construct the disjunctive and conjunctive lattice PWA approximations,  only $8^4=4096$ samples are generated uniformly in the region $\Omega =[-0.6~0.6]\times [-0.9~0.9]\times [-0.21~0.21]\times [-0.6~0.6]$. In this case, there are only 13 distinct affine functions, which is due to the choice of the region. For a larger region, say $\Omega _2= [-0.9~0.9]\times [-1.2~1.2]\times [-0.25~0.25]\times [-0.9~0.9]$, there are 47 distinct affine functions. However, there are a significant number of infeasible state points in $\Omega_2$, and according to the state trajectory in \cite{Karg2020efficient}, the region $\Omega$ is enough. 

The evaluation of Algorithm \ref{alg:sampling}-\ref{alg:addpoints} results in the disjunctive and conjunctive lattice PWA approximations, both with six terms. All the computations in this paper are implemented through MatLab 2016b (MathWorks, USA) on an Apple M1 Max computer.

\cbstart The entire offline calculation time is 41.25s, i.e., Algorithm 1, 2, 3 takes 40.25s, 0.66s and 0.33s, respectively.
The number of parameters for both approximations is 143, while the number of parameters stored in the MPT3 solution is 104,535. The average online evaluation time for the MPT solution, the disjunctive as well as conjunctive lattice PWA approximations are 0.0052s, $8.96\times 10^{-5}$s, and $8.33 \times 10^{-5}$s, respectively. It is apparent that the approximation results in a much lower online computational burden.\cbend
 
It has been tested through $5\times 10^6$ test data points that the two lattice PWA approximations are identical in the  region $\Omega$. By setting $\epsilon=10^{-3}$, it can be concluded that with confidence $\delta =0.9999$ the probability that the approximated lattice PWA control laws equal the optimal control is larger than 0.999. For the $5\times 10^6$ points, the optimal explicit linear MPC control law is also calculated and it is found  that the lattice PWA approximations are error-free in $\Omega$.

Fig. \ref{ex2_fig} shows one exemplary closed-loop simulation of the example, and we can see from the figure that the optimal state trajectory and the trajectory with the lattice PWA approximations as inputs are identical.
\begin{figure}[htbp]
\centering
 \psfrag{x1}[c]{$\Phi$}
  \psfrag{u}[c]{$u$}
   \subfigure[]{
  \includegraphics[width=0.8\columnwidth]{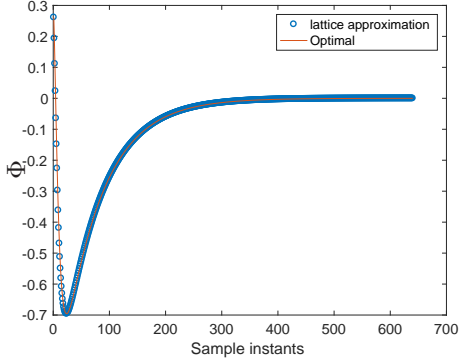}}
  \quad
  \subfigure[]{
    \includegraphics[width=0.8\columnwidth]{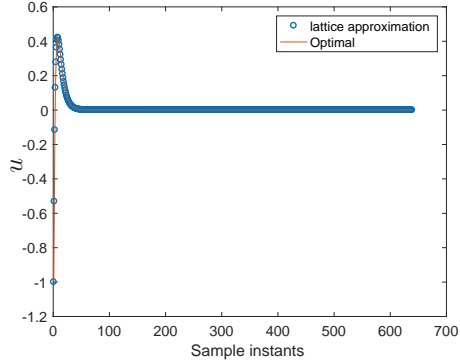}}
    \caption{One exemplary closed-loop simulation of Example \ref{example2}.}
  \label{ex2_fig}
\end{figure}

\end{exmp}

\begin{exmp}\label{example3}
\cbstart
Consider an example taken from \cite{Moshkenani2018combinatorial}, which is an mpQP problem constructed from the typical MPC setup of the form
\begin{subequations}
\begin{align}
&\min\limits_{u_0, \ldots, u_{N-1}}\bm x_N^TP\bm x_N+\sum\limits_{k=0}^{N-1}\bm x_k^TQ\bm x_k+u_k^TRu_k\\
&~~~~s.t.\bm x_{k+1}=A\bm x_k+Bu_k\\
&~~~~~~~~\bm x \in \mathcal{X}, u \in \mathcal{U}
\end{align}
\end{subequations}
with $\bm x \in \mathbb{R}^{10}$, $u \in \mathbb{R}$, $P=Q=I_{10}, R=1$, $\mathcal{X}=\{x|-10 \leq x_i \leq 10, i=1, \ldots, n_x\}$, $\mathcal{U}=\{-1\leq u \leq 1\}$. The prediction model is obtained by discretizing the model $1/(s+1)^{10}$ with sampling time of 1 $\mathrm{s}$ and then converting the discretized model into state space form. 
The prediction horizon is taken to be $N=10$. 

Apparently this problem is more complex than Example \ref{example2}, as the dimension is much higher. As \cite{Moshkenani2018combinatorial} shows, the $mpt_{solve}$ function in MPT3 Toolbox failed to solve this problem offline, i.e., MATLAB ran out of memory.

 Here, 
in order to construct the disjunctive and conjunctive lattice PWA approximations,  we have to generate sample points in the 10-dimensional domain, which may suffer from the curse-of-dimensionality if the sample points are generated uniformly in the domain. Instead, we generate sample points on several trajectories, which reflect the most possible states of the dynamic system under optimal MPC control law. Specifically, 300 initial points are generated in the domain $[-2, 2]^{10}$, then for the 300 trajectories we have 7802 sample points.


 In this case, there are 16 distinct affine functions. The evaluation of Algorithm \ref{alg:sampling}-\ref{alg:addpoints} results in the disjunctive and conjunctive lattice PWA approximations, the number of terms is 9 and 8, respectively. The offline as well as online complexity of lattice PWA approximations when the prediction horizon $N$ is 10 are shown in Table \ref{tab:result_ex3}. 
 \begin{table}[htbp]
\caption{Performance of lattice PWA approximations on Example \ref{example3}, in which $N_s$ and $\#\rm{para}$ are the number of  sample points and parameters, and $t_{\rm off}$ and $t_{\rm on}$ denote offline and online computation time.}
\label{tab:result_ex3}
\begin{center}
\begin{tabular}{cccccc}
\hline
Method&$N$&$N_s$&\#para&$t_{\rm off}$[s]&$t_{\rm on}$[s]\\
Disjunctive&10&7808&320&84.65&$1.69\times 10^{-4}$\\
Conjunctive&&7808&304&84.65&$1.40 \times 10^{-4}$\\
Online MPC&&&&&0.0077\\
Method in \cite{Moshkenani2018combinatorial}&&&&72.85&\\
\hline
Disjunctive&20&10188&418&229.24&$2.71\times 10^{-4}$\\
Conjunctive&&10188&418&229.24&$2.53 \times 10^{-4}$\\
Online MPC&&&&&0.015\\
Method in \cite{Moshkenani2018combinatorial}&&&&&\\
\hline
\end{tabular}
\end{center}
\end{table}

The entire offline calculation time for generating the disjunctive and conjunctive lattice PWA approximations is 84.65s, i.e., Algorithm 1,2,3 accounts for 80.3016s, 2.8734s, 1.4776s, respectively.
The number of parameters for the disjunctive and conjunctive approximations are 320 and 304, respectively. The average online evaluation times (over $10^4$ trials) for the online QP, disjunctive as well as conjunctive lattice PWA approximations are $0.0077$, $1.69\times 10^{-4}$s, and $1.40 \times 10^{-4}$s, respectively. It is apparent that the  approximations result in a very low online computational burden.

It has been tested through $5\times 10^6$ test data points that the two lattice PWA approximations are identical in the  region $\Omega$. By setting $\epsilon=10^{-3}$, it can be concluded that with confidence $\delta =0.9999$ the probability that the approximated lattice PWA control laws equal the optimal control law is larger than 0.999. For the $5\times 10^6$ points, the optimal linear MPC control law is also calculated and it is found  that the lattice PWA approximations are error-free in $\Omega$.

 Fig. \ref{ex3_fig} shows one exemplary closed-loop simulation of the example, and we can see from the figure that the optimal state trajectory and the trajectory with the lattice PWA approximations as inputs are identical.

 \begin{figure}[htbp]
 \centering
  \psfrag{x1}[c]{$x_1$}
   \psfrag{u}[c]{$u$}
    \subfigure[]{
   \includegraphics[width=0.8\columnwidth]{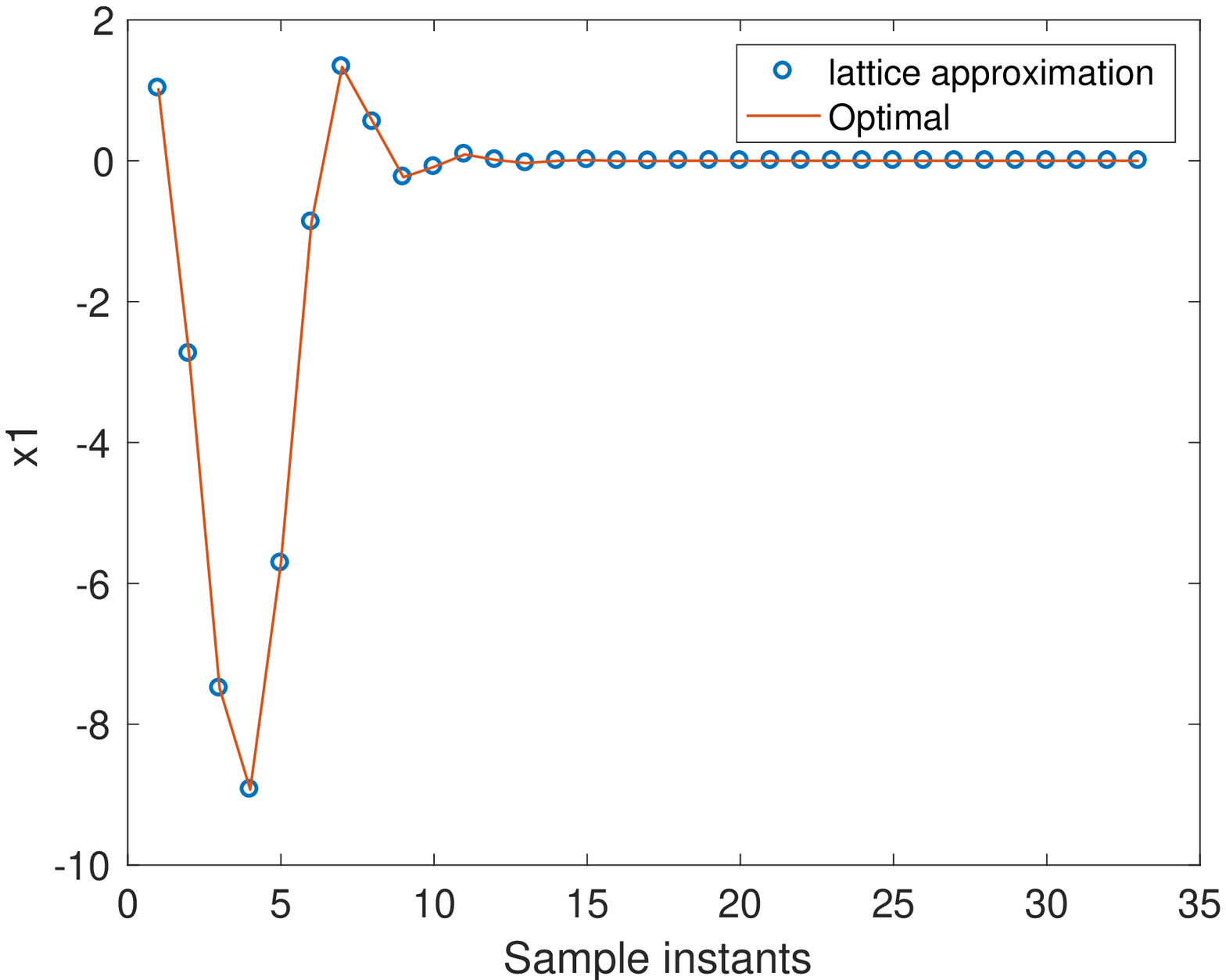}}
   \quad
   \subfigure[]{
     \includegraphics[width=0.8\columnwidth]{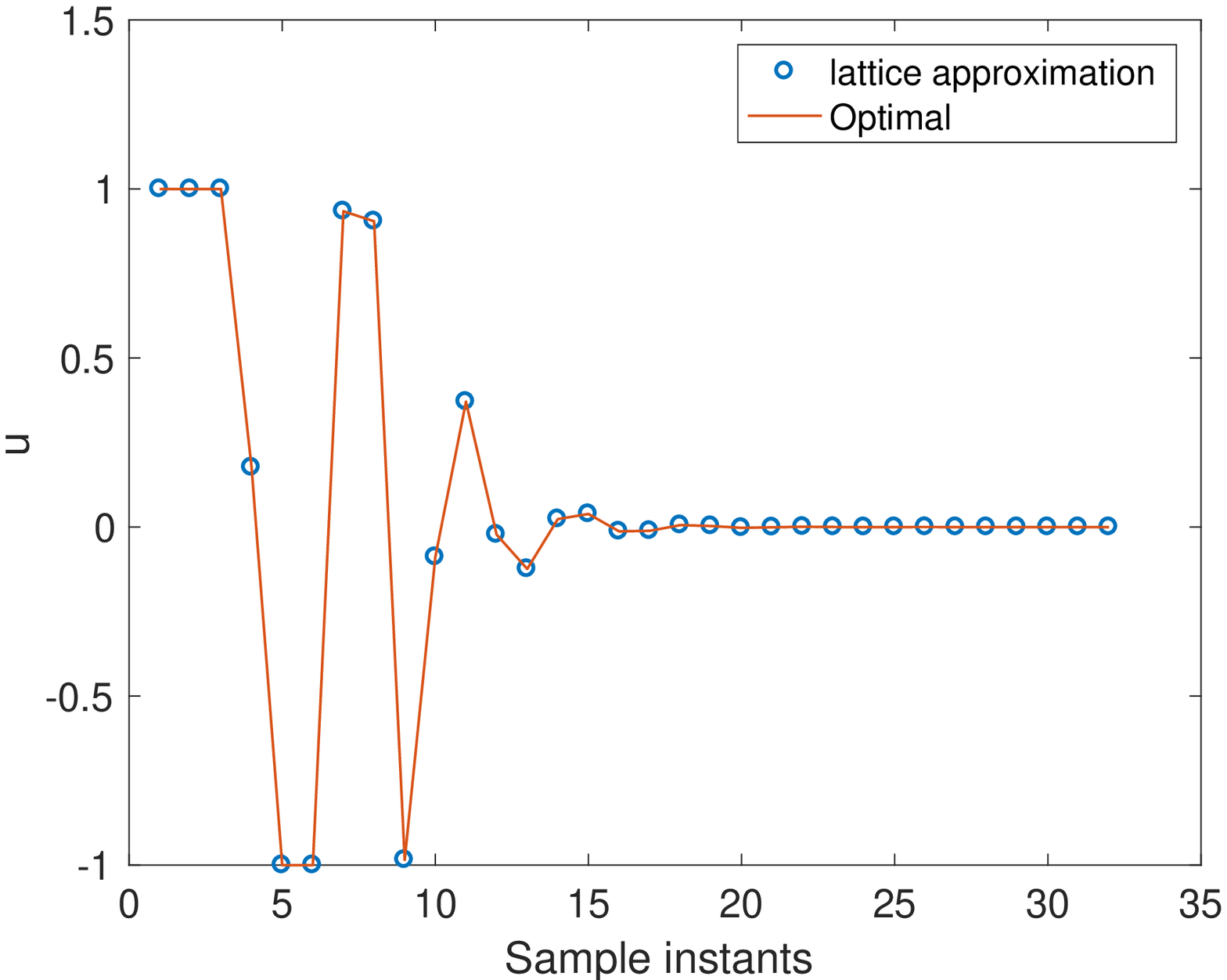}}
     \caption{One exemplary closed-loop simulation of Example \ref{example3}.}
   \label{ex3_fig}
 \end{figure}


As indicated in \cite{Moshkenani2018combinatorial}, the MPT3 Toolbox failed to solve the problem. Moreover the combinatorial approach proposed in \cite{Moshkenani2018combinatorial} successfully enumerating all optimal active sets and creating corresponding critical regions, takes 72.85s. It is noticed that although the calculation time for our procedure is slightly longer, we obtain the final continuous PWA expression of the optimal control law in a simplified form, which  is easier to be implement online. 

To demonstrate more clearly the efficacy of the lattice PWA approximations, the prediction horizon $N$ is extended to 20, and the complexity of lattice PWA approximations is also listed in Table \ref{tab:result_ex3}. Besides, we also generate $5\times 10^6$ validation points to show that with confidence $\delta =0.9999$ the probability that the approximated lattice PWA control laws equal the optimal control law is larger than 0.999. When $N=20$, \cite{Moshkenani2018combinatorial} did not provide a result either. The lattice PWA approximations in this case takes 229.2417s to get error-free approximations in $\Omega=[-2, 2]^{10}$, showing that the lattice PWA approximations scale well with the problem size for this example. 
\cbend
\end{exmp}

\section{Conclusions and Future work}
In this paper, we have presented disjunctive and conjunctive lattice PWA approximations of the explicit linear MPC control law by sampling and resampling in the state domain. The lattice PWA approximated and exact control laws are identical for sample points and in UO regions that contain the sample points as interior points. Furthermore, under the assumption that all the affine functions have been identified in the domain of interest, the disjunctive lattice PWA approximation is always smaller than the original optimal control law, while the conjunctive lattice PWA approximation is always larger. Then if the disjunctive and conjunctive lattice PWA approximations are identical, both are equivalent to the optimal control law. The two kinds of lattice PWA approximations have been simplified to further reduce the storage and online evaluation complexity. The complexity of the online and offline approximation as well as the storage requirements, have been analyzed. Simulation results show that with a moderate number of sample points we can obtain statistically error-free lattice PWA approximations that are calculated with relatively small computational cost.

\cbstart
In the future, apart form the information of local affine functions, the information of critical regions will also be used to construct the lattice PWA approximation. Moreover, for more general  domain of interest, the approximation error will be considered, which can be used to derive corresponding feasibility and stability analysis. Besides, practical applications will also be considered.
\cbend


\bibliographystyle{unsrt}        
\bibliography{/Users/Jun/Documents/paper_work/refs.bib}           



\end{document}